\documentclass[a4paper,11pt]{article}
\pdfoutput=1 

\usepackage{jheppub} 

\usepackage[T1]{fontenc} 
\usepackage[textsize=footnotesize,textwidth=2.5cm]{todo notes}

\definecolor{mikadoyellow}{rgb} {0.16, 0.254, 0.6}

\usepackage{amssymb,amsmath,latexsym,bm,amsfonts}
\usepackage{graphicx}
\usepackage{longtable}
\usepackage{color,xcolor}
\usepackage{indentfirst}
\usepackage{subfigure}
 \usepackage{comment}

\newcommand{\bra}[1]{\ensuremath{\left\langle#1\right|}}
\newcommand{\ket}[1]{\ensuremath{\left|#1\right\rangle}}

\newcommand{\be}{\begin{equation}}
\newcommand{\ee}{\end{equation}}
\newcommand{\bpm}{\begin{pmatrix}}
\newcommand{\epm}{\end{pmatrix}}

\newcommand{\beqn}{\begin{eqnarray}}
\newcommand{\eeqn}{\end{eqnarray}}

\newcommand{\besub}{\begin{subequations}}
\newcommand{\eesub}{\end{subequations}}
\newcommand{\bea}{\begin{eqnarray}}
\newcommand{\eea}{\end{eqnarray}}

\title{ \boldmath Balanced Partial Entanglement and the Entanglement Wedge Cross Section}

\author[a,b]{Qiang Wen}

\affiliation[a]{Shing-Tung Yau Center of Southeast University, Nanjing 210096, China}
\affiliation[b]{School of Mathematics, Southeast University, Nanjing 211189, China}

\emailAdd{wenqiang@seu.edu.cn}

\abstract{In this article we define a new information theoretical quantity for any bipartite mixed state $\rho_{AB}$. We call it the \textit{balanced partial entanglement} (BPE). The BPE is the partial entanglement entropy, which is an integral of the entanglement contour in a subregion, that satisfies certain balance requirements. The BPE depends on the purification hence is not intrinsic. However, the BPE could be a useful way to classify the purifications. We discuss the entropy relations satisfied by BPE and find they are quite similar to those satisfied by the entanglement of purification. We show that in holographic CFT$_2$ the BPE equals to the area of the entanglement wedge cross section (EWCS) divided by 4G. More interestingly, when we consider the canonical purification the BPE is just half of the reflected entropy, which also directly relate to the EWCS. The BPE can be considered as an generalization of the reflected entropy for a generic purification of the mixed state $\rho_{AB}$. We interpret the correspondence between the BPE and EWCS using the holographic picture of the entanglement contour.}

\begin{document} 
\maketitle
\flushbottom

\section{Introduction}
Quantum entanglement has played a central role in the study of modern theoretical physics, including quantum information theory, condensed matter theory and quantum gravity. One of the most important entanglement measures is the entanglement entropy, which captures the correlation between $A$ and $B$ for a bipartite system $A\cup B$ in a pure state. The study of entanglement entropy gains huge amount of extra attention because of the Ryu-Takayanagi (RT) \cite{Ryu:2006bv,Ryu:2006ef} proposal, which reveals the deep connection between spacetime geometry and quantum entanglement. In the context of the AdS/CFT correspondence \cite{Maldacena:1997re,Gubser:1998bc,Witten:1998qj}, let us consider a static region $A$ in the boundary field theory and the minimal surface $\mathcal{E}_{A}$ in the dual AdS bulk that anchored on the boundary of $A$. The RT proposal relates the entanglement entropy of $A$ to the area of $\mathcal{E}_{A}$ in Planck units, i.e.
\begin{align}
S_{A}=\frac{1}{4G}\text{Area}(\mathcal{E}_{A})\,.
\end{align}

However, for bipartite (or multipartite) systems in a mixed state, the entanglement entropy is not a good measure of entanglement as it mixes classical and quantum correlations. New entanglement measures in quantum information theory are proposed to take the place, for example, the mutual information, the (logarithmic) entanglement negativity \cite{negativity1,negativity2,negativity3}, the entanglement of purification (EoP) \cite{EoP}, the partial entanglement entropy (PEE) \cite{Vidal,Wen:2018whg,Kudler-Flam:2019oru,Wen:2019ubu,Wen:2019iyq} etc. 

On the other hand, from the perspective of holography there exists a geometric quantity called the entanglement wedge cross section (EWCS) $\Sigma_{AB}$, that may capture certain type of correlations between $A$ and $B$ for a mixed state $\rho_{AB}$. For example, consider $AB\equiv A\cup B$ to be a subsystem of the boundary of global AdS$_3$, the reduced density matrix $\rho_{AB}$ has a bulk dual named the entanglement wedge $\mathcal{W}_{AB}$ \cite{Czech:2012bh,Wall:2012uf,Headrick:2014cta}. The entanglement wedge is the causal development of the homology surface $\mathcal{R}_{AB}$, which is a Cauchy surface with the boundary being $\partial {\mathcal{R}_{AB}}=A\cup B\cup \mathcal{E}_{AB}$. The EWCS $\Sigma_{AB}$ is then defined as the minimal cross section of $\mathcal{R}_{AB}$ that separate $A$ from $B$. Since $\Sigma_{AB}$ plays a special role in the bulk, it is very likely to represent something special in quantum information theory.

So far, there are several proposals for the holographic dual of $\Sigma_{AB}$. The first candidate is the entanglement of purification (EoP) \cite{EoP}. It was shown in \cite{HEoP1,HEoP2} that the EWCS and EoP satisfy the same entropy relations in holographic theories. Also assuming a theory with a tensor network description thus the \textit{\textit{surface/state correspondence}} \cite{Miyaji:2015yva} can be realized, the calculation of $\Sigma_{AB}$  matches with the way we define the EoP. However, it could be very hard to justify this proposal in more generic cases due to the large optimization procedure in the definition of the EoP. Another well-known candidate is half of the reflected entropy \cite{Dutta:2019gen}, which is defined on the canonical purification of $\rho_{AB}$. The reflected entropy proposal can be confirmed under some mild assumptions. There are also other proposals which claim that the EWCS is dual to, for example the logarithmic negativity \cite{Kudler-Flam:2018qjo,Kusuki:2019zsp}, the  ``odd entropy'' \cite{Tamaoka:2018ned}, the ``differential purification'' \cite{Espindola:2018ozt} and so on. See also \cite{Agon:2018lwq,Harper:2019lff,Bao:2019wcf,Lin:2020yzf,Du:2019emy,Umemoto:2018jpc,BabaeiVelni:2019pkw,Ghodrati:2019hnn,Ghodrati:2020vzm,KumarBasak:2021lwm,Basak:2020oaf,Chu:2019etd,Khoeini-Moghaddam:2020ymm} for a incomplete list about recent studies related to the EWCS. It is worth pointing out that the above proposals are indeed in tension with each other (see for example \cite{Akers:2019gcv}), hence a deeper understanding of the holographic picture for the EWCS is still in need. 

Recently a new entanglement measure called the partial entanglement entropy (PEE) \cite{Vidal,Wen:2018whg,Kudler-Flam:2019oru,Wen:2019ubu,Wen:2019iyq} was proposed. For a given region $A$ and a subset $A_i$ of $A$, the PEE is denoted as $s_{A}(A_i)$. Physically it is assumed to capture the contribution from $A_i$ to the entanglement entropy $S_{A}$. The key property featured by the PEE is additivity, which is not possessed by any other entanglement measures. The differential version of the PEE, named the entanglement contour \cite{Vidal}, is a function $f_{A}(\textbf{x})$ defined on $A$ that gives the contribution from the degrees of freedom at any point $\textbf{x}$ in $A$ to $S_A$. In other words it is the density function of the entanglement entropy $S_A$ that satisfies,
\begin{align}
S_{A}=\int_{ A}f_{A}(\textbf{x})dx^d\,.
\end{align}
Here $d$ gives the dimension of $A$. The PEE $s_{A}(A_i)$ is then defined in the following
\begin{align}\label{a2definition}
s_{A}(A_i)=\int_{ A_i}f_{A}(\textbf{x})dx^d\,.
\end{align}
Note that the PEE $s_{A}(A_i)$ only collect the contribution in the subset $A_i$.

Let us assume that $\bar{A}$ is some system that purifies $A$.  Since the  PEE $s_{A}(A_i)$ in some sense captures the correlation between the subset $A_i$ and $\bar{A}$, it should be invariant under the permutation between $A_i$ and $\bar{A}$ \cite{Wen:2019iyq}. In order to manifest this permutation symmetry, we also write the PEE in the following way
\begin{align}
s_{A}(A_i)=\mathcal{I}(A_i,\bar{A})=\mathcal{I}(\bar{A},A_i)=s_{\bar{A}_i}(\bar{A}) \equiv\mathcal{I}_{A_i \bar{A}}.
\end{align}
 Note that we should not mix the PEE $\mathcal{I}(A_i,\bar{A})$ with the mutual information $I(A_i,\bar{A})$.

Unfortunately, the fundamental definition based on the reduced density matrix for the PEE is still missing. According to its physical meaning, the PEE should satisfy the following physical requirements \footnote{ The requirements 1-6 are firstly given in \cite{Vidal}, while the requirement 7 is recently given in \cite{Wen:2019iyq}}:
\begin{enumerate}
\item \textbf{Additivity}: if $A_{i}^{a}\cup A_{i}^{b}=A_{i}$ and $A_{i}^{a}\cap A_{i}^{b}=\varnothing$, by definition we should have
\begin{align}\label{additivity}
s_{A}(A_{i})&=s_{A}(A_{i}^{a})+s_{A}(A_{i}^{b})\,.
\end{align}

\item \textbf{Invariance under local unitary transformations}: $s_{A}(A_i)$ should be invariant under any local unitary transformations inside $A_i$ or $\bar{A}$.

\item \textbf{Symmetry}: for any symmetry transformation $\mathcal{T}$ under which $\mathcal{T}A=A'$ and  $\mathcal{T} A_{i}= A'_{i}$, we have $s_{A}(A_i)=s_{A'}(A'_i)$.

\item \textbf{Normalization}: $ S_{A}=s_{A}(A_{i})|_{A_i\to A}\,.$

\item \textbf{Positivity}: $s_{A}(A_i)\ge 0$.

\item \textbf{Upper bound}: $
 s_{A}(A_i) \leq S_{A_{i}} \,.
$

\item \textbf{Symmetry under the permutation}: $\mathcal{I}(\bar{A},A_i)=\mathcal{I}(A_i,\bar{A})\,,$ which implies $s_{A}(A_i)=s_{\bar{A}_i}(\bar{A})$.
\end{enumerate}

Recent explorations on entanglement contour or the PEE include \cite{Vidal:2002rm,Botero,Vidal,PhysRevB.92.115129,Coser:2017dtb,Tonni:2017jom,Alba:2018ime,Wen:2018whg,Wen:2018mev,Kudler-Flam:2019oru,Wen:2019ubu,DiGiulio:2019lpb,Han:2019scu,Ageev:2019fjf,Kudler-Flam:2019nhr,Wen:2019iyq,Roy:2019gbi,deBuruaga:2019xwv,MacCormack:2020auw}. People propose formulas to construct the PEE (or entanglement contour) that satisfies the above requirements. Each of the existed proposals are restricted to special configurations. The first one is the Gaussian formula \cite{Botero,Vidal,PhysRevB.92.115129,Coser:2017dtb,Tonni:2017jom,Alba:2018ime,DiGiulio:2019lpb,Kudler-Flam:2019nhr} that applies to the Gaussian states in free theories, where the density matrix can be completely characterized in terms of the correlation matrix. The second proposal is a geometric construction \cite{Wen:2018whg,Wen:2018mev,Han:2019scu} in holograhic theories, which is inspired by a natural slicing of the entanglement wedge following the boundary and bulk modular flows.  The third one is the partial entanglement entropy proposal \cite{Wen:2018whg,Wen:2019ubu} that claims the PEE is given by an additive linear combination of subset entanglement entropies. The fourth proposal \cite{Wen:2019iyq} follows the construction of the extensive (or additive) mutual information (EMI) \cite{Casini:2008wt} (see also \cite{Roy:2019gbi} for a related construction), which tried to solve the above seven requirements in CFT. 

Though the above proposals have very different physical motivations, the PEE calculated by different approaches are highly consistent with each other \cite{Kudler-Flam:2019nhr, Wen:2018whg,Wen:2018mev,Han:2019scu,Wen:2019iyq}. This implies the PEE should be unique and well defined. So far, the uniqueness of the PEE is only confirmed for Poincar\'e invariant theories  \cite{Wen:2019iyq}, by showing that the above seven requirements in these theories have unique solution. Due to the nice properties, the PEE is also useful to study the entanglement structure in condensed matter theories
\footnote{The entanglement contour gives a finer description for the entanglement structure. In condense matter theories it can be used to discriminate between gapped systems and gapless systems with a finite number of zero modes in $d=3$ \cite{Vidal}. It has been shown to be particularly useful to characterize the spreading of entanglement when studying dynamical situations \cite{Vidal,Kudler-Flam:2019oru,DiGiulio:2019lpb}. Modular flows in two dimensions can be generated from the PEE \cite{Wen:2019ubu}. The entanglement contour is also a useful probe of slowly scrambling and non-thermalizing dynamics for some interacting many-body systems \cite{MacCormack:2020auw}. Holographically the PEE \cite{Wen:2018whg,Wen:2018mev} correspond to bulk geodesic chords which is a finer correspondence between quantum entanglement and bulk geometry \cite{Wen:2018mev,Abt:2018ywl}. The new concept of entanglement contour in quantum information will play an important role in our understanding of the gauge/gravity duality and the entanglement structure in quantum field theories (or many-body system).}.

Since the entanglement contour is a finer description for the entanglement structure, we expect that other entanglement measures could be extracted from the contour function. In this paper, we study the correlations in mixed bipartite states using PEE. For a mixed state $\rho_{AB}$, one can introduce an auxiliary system $A'B'$ thus the combined system is in a pure state $\ket{\psi}$, which is highly non-unique. The state $\ket{\psi}$ is then called a purification of $\rho_{AB}$. In section \ref{section2}, we briefly review the PEE proposal and define a special PEE for any purification. We call it the \textit{balanced partial entanglement} (BPE), because the partition of $A'B'$ should satisfy the following balance requirement, $s_{AA'}(A)=s_{BB'}(B)$. In section \ref{section3}, we study aspects of the BPE for the case where $A$ and $B$ are adjacent. We calculate the holographic BPE for the case of global AdS$_{3}$, where $AB$ is a subsystem on the boundary. We find the BPE gives the area of the EWCS. Interestingly we find the crossing PEE $\mathcal{I}_{AB'}=\frac{c}{6}\log 2$, which is a constant independent from the length of $A$ and $B$. We discuss the entropy relations satisfied by the BPE and find that, they are quite similar to those satisfied by EoP. Also, we consider the minimal purification in the context of the \textit{surface/state correspondence} \cite{Miyaji:2015yva}, and find the BPE differs from the case of the global AdS$_3$. In section \ref{section4}, we discuss the cases where $A$ and $B$ are non-adjacent. We confirmed that the relation between the BPE and $\Sigma_{AB}$ still holds. In section \ref{section5}, we discuss the canonical purification for generic $\rho_{AB}$ and show that, half of the reflected entropy is identical to the BPE we defined. In section \ref{section6}, we interpret the relation between the BPE and the EWCS using the holographic picture for PEE, which are the geodesic chords in the bulk that normal to the RT surfaces of relevant regions. At last we give a discussion in section \ref{sectiond}.

\section{The balanced partial entanglement}\label{section2}

\subsection{Definition}

\begin{figure}[h] 
   \centering
   \includegraphics[width=0.45\textwidth]{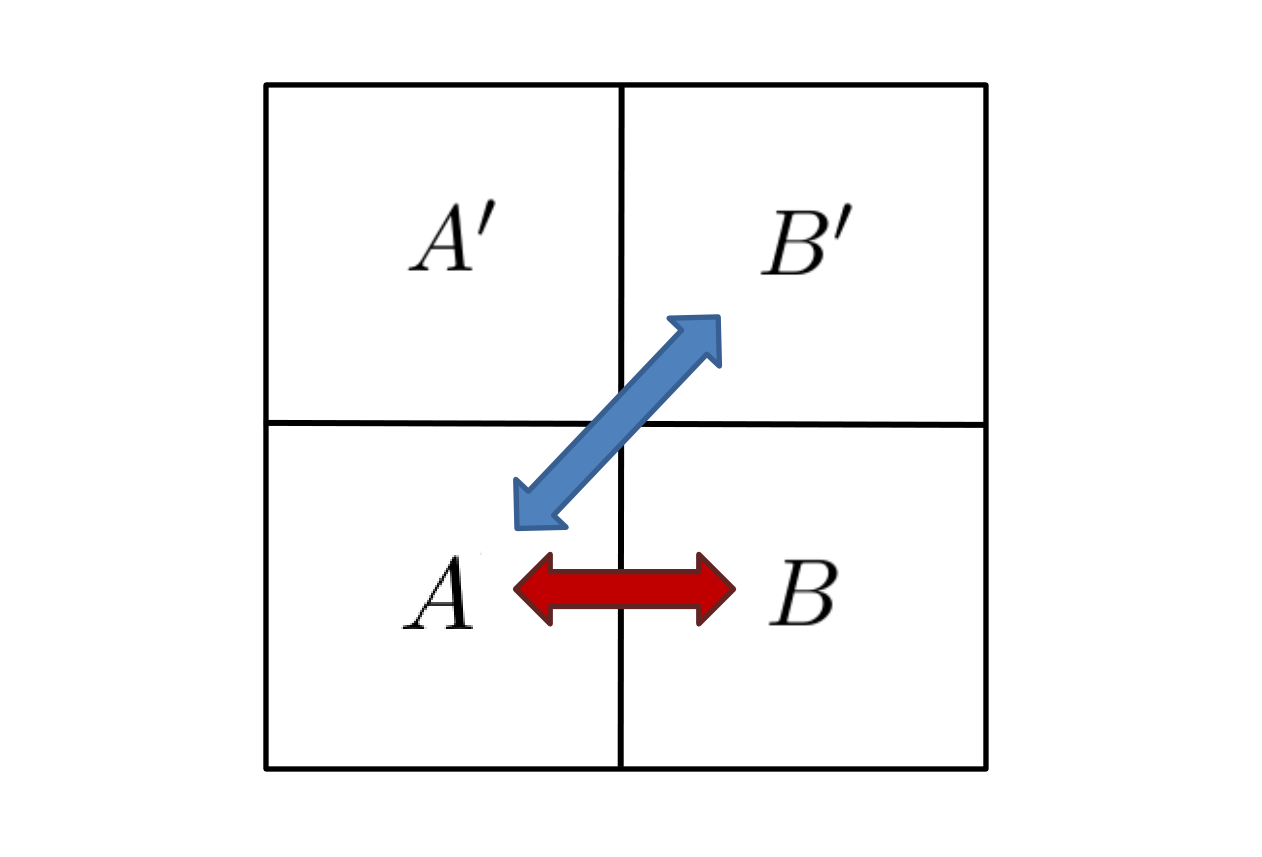}  
 \caption{The combined system $ABA'B'$ is in a pure state. The red arrow represents the PEE $\mathcal{I}_{AB}$ between $A$ and $B$, while the blue arrow represents $\mathcal{I}_{AB'}$. $\mathcal{I}_{AB'}$ and $\mathcal{I}_{BA'}$ vary with the partition of the complement $A'B'$. The balance requirement $\mathcal{I}_{AB'}=\mathcal{I}_{BA'}$ can be satisfied by adjusting the partition.
\label{ABApBp} }
\end{figure}

The \textit{balanced partial entanglement} (BPE) is defined in the following: let $\rho_{AB}$ be a density matrix on a bipartite system $\mathcal{H}_{A}\otimes \mathcal{H}_{B}$. We consider an auxiliary system $A'B'$ that purifies $AB$, thus the whole system is in a pure state $\ket{\psi}$ and $\text{Tr}_{A'B'}\ket{\psi}\bra{\psi}=\rho_{AB}$. Let us partition the auxiliary system into $A'$ and $B'$ properly in the following way. Firstly we require the contribution from $A$ to the entanglement entropy $S_{AA'}$ equals to the contribution from $B$ to the entanglement entropy $S_{BB'}$. We call this requirement the \textit{balance requirement}, i.e. 
\begin{align}\label{requirement1}
\textit{balance requirement}:\qquad s_{AA'}(A)=s_{BB'}(B)\,,\qquad s_{AA'}(A')=s_{BB'}(B')\,.
\end{align}
Since $S_{AA'}=S_{BB'}$, only one of the above requirements is independent. In terms of the PEE we have,
\begin{align}
s_{AA'}(A)=\mathcal{I}_{AB}+\mathcal{I}_{AB'}\,,\qquad s_{BB'}(B)=\mathcal{I}_{BA}+\mathcal{I}_{BA'}\,.
\end{align}
See Fig.\ref{ABApBp}. The first term $\mathcal{I}_{AB}$ is supposed (though not proved in general) to be intrinsic hence independent from the purification, because unitary transformations acting outside $AB$ should not change the correlation between $A$ and $B$. While the second term $\mathcal{I}_{AB'}$ could vary under different purifications. Since $\mathcal{I}_{AB}=\mathcal{I}_{BA}$, the balance requirement can also be written as
\begin{align}\label{requirement1a}
\mathcal{I}_{AB'}=\mathcal{I}_{BA'}\,.
\end{align}
When $\mathcal{I}_{AB'}$ (or $\mathcal{I}_{BA'}$) satisfies the balance requirement, we call it the \textit{crossing PEE} of $\ket{\psi}$, which can be used to classify the purifications.

The BPE is a natural quantity to consider for any purification $\ket{\psi}$ of $\rho_{AB}$. However, in general the partition of $A'B'$ that satisfies the balance requirement is not unique. To clarify this ambiguity we propose the following minimal requirement.
\begin{itemize}
\item \textit{Minimal requirement}:  among all the partitions that satisfy the balance requirement we should choose the one such that $s_{AA'}(A)$ reaches its minimal value.
\end{itemize}
 It is not hard to find a solution to this requirement since the purification is already fixed. In most theories, the entanglement between any two local degrees of freedom decreases with distance. 
When we set $B'$ to be as far from $A$ as we can and set $A'$ to be as far from $B$ as we can, we can simultaneously reduce $\mathcal{I}_{AB'}$ and $\mathcal{I}_{BA'}$ while keeping the balance requirement satisfied. Eventually we arrive at the following prescription for the natural and simple partition: The whole system is partitioned into two parts $AA'$ and $BB'$ thus separates $A$ from $B$. In some sense the region $A$ will be surrounded by $A'$ while $B$ is surrounded by $B'$. Also, there should be no embedding between the region $B'$ and $A'$. Later we will partition the purifier $A'B'$ following this prescription and \textit{not stress the minimal requirement any more}.

\begin{figure}[h] 
   \centering
   \includegraphics[width=0.325\textwidth]{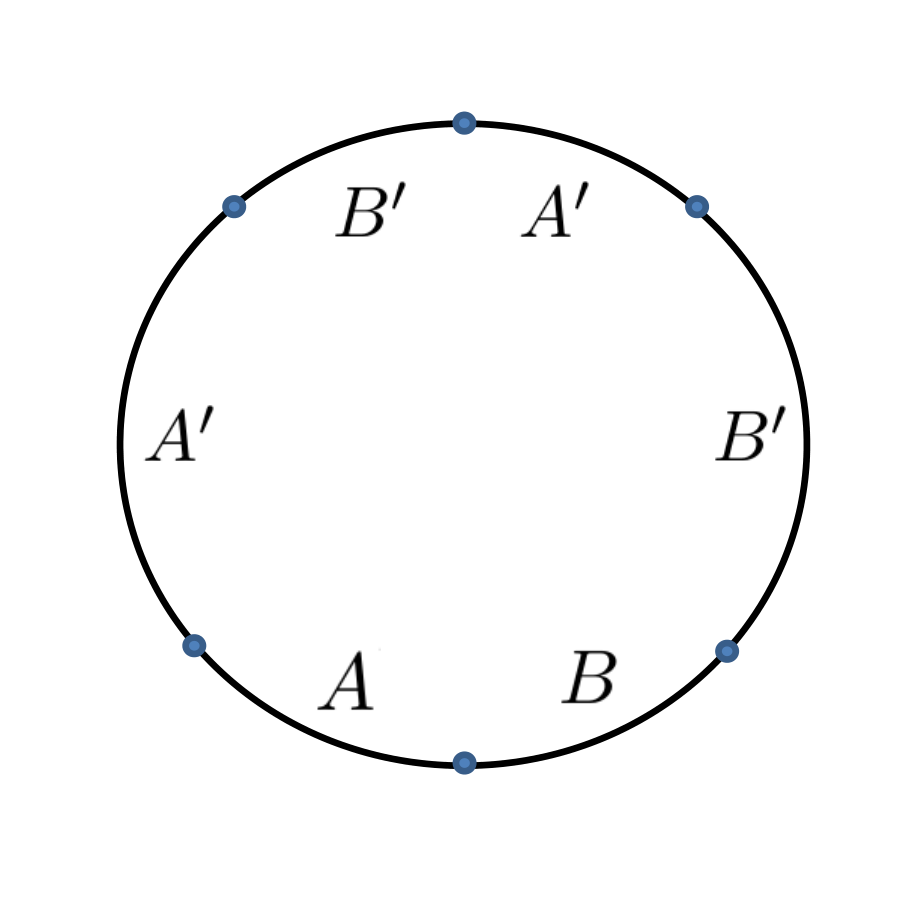}  \includegraphics[width=0.325\textwidth]{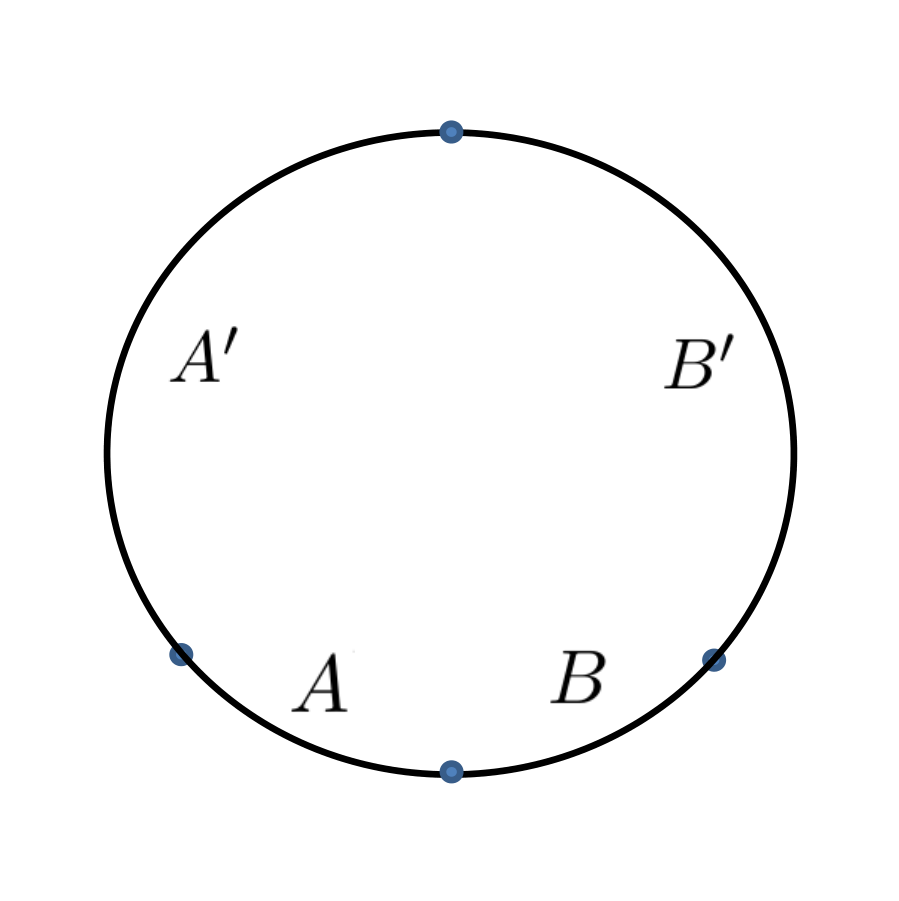} \includegraphics[width=0.325\textwidth]{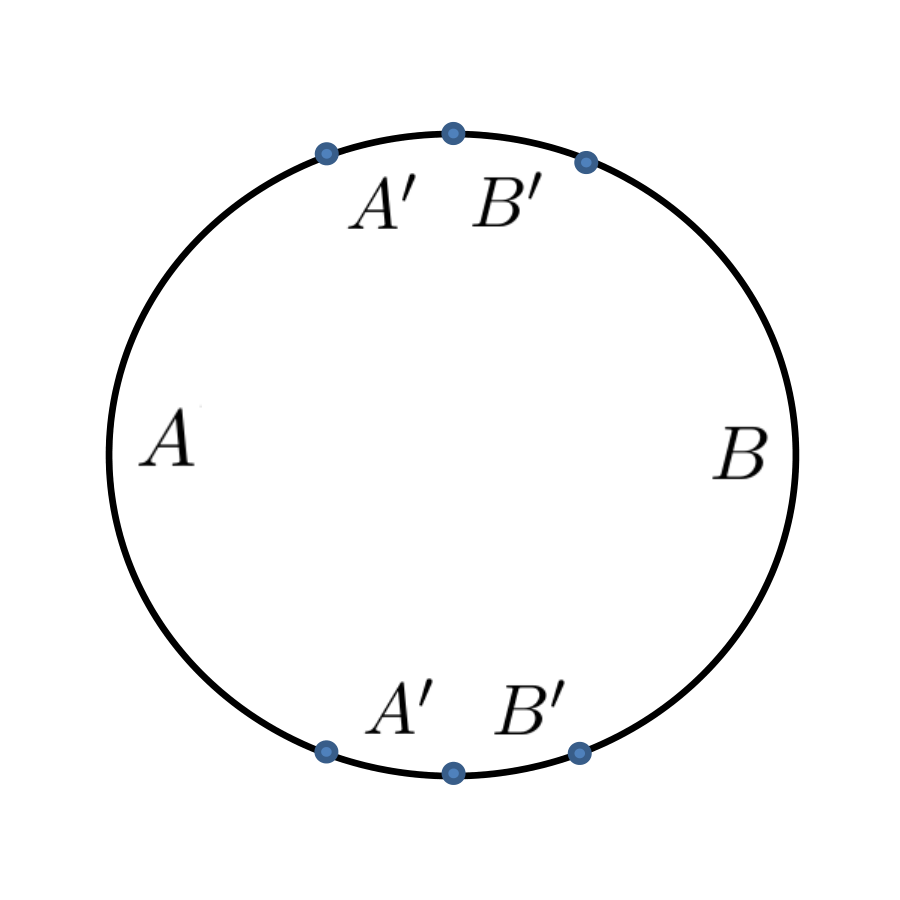}
 \caption{All the three figures have reflection symmetry between the left and right hand side, hence we can easily see that the balance requirement is satisfied in all the figures. From the first figure to the second one, the upper part of $A'$ and $B'$ exchange positions hence the upper part of $B'$ gets further from $A$ thus reduces the PEE between them. The third figure shows the proper partition that satisfies our requirements for the case where $AB$ is non-adjacent.
\label{partitions} }
\end{figure}

One can understand our prescription via the examples in Fig.\ref{partitions}. In the first two figures $AB$ is a connected interval in a circle, while the partition of the complement $A'B'$ is different. The circle is in a pure state. We require these two configurations to have reflection symmetry between the left and right hand side, hence the balance requirement can be satisfied in both cases. It is obvious that the $s_{AA'}(A)$ reaches its minimal value in the second figure because, compare with the first figure, part of $B'$ get further from $A$ hence $\mathcal{I}_{AB'}$ decreases.

Then the BPE is defined as the partial entanglement entropy $s_{AA'}(A)$ under the partition of $A'B'$ satisfying the above two requirements. We denote it as BPE$(A,B,\psi)$, i.e.
\begin{align}
\text{BPE}(A,B,\psi)=s_{AA'}(A)|_{balance~requirements}\,.
\end{align}
 When the purification is specified we will omit the label $\psi$ thus write BPE$(A,B)$.

The third figure in Fig.\ref{partitions} shows the proper partition for the reflection symmetric case where $AB$ is non-adjacent. For more generic cases with no reflection symmetry, the boundary is partitioned similarly by two points and the both of the region $A'=A'_1\cup A'_2$ and $B'=B'_1\cup B'_2$ contain two disconnected pieces which can be naturally set in pairs. For example in Fig.\ref{nonadjacent}, the two pairs are $A_1'\sim B_1'$ and $A_2'\sim B_2'$. Note that in the same sense $A\sim B$ are also a pair. The balance requirements are indeed imposed on all the pairs,
\begin{align}\label{BR2}
s_{AA'}(A_1')=s_{BB'}(B_1')\,,\qquad s_{AA'}(A)=s_{BB'}(B)\,,
 \qquad s_{AA'}(A_2')=s_{BB'}(B_2')\,.
\end{align}
Since $S_{AA'}=S_{BB'}$, only two of the above requirements are independent, which are exactly the requirements that determine the two partition points.

\subsection{Review on the partial entanglement entropy proposal}
In order to study the BPE, we need to calculate the PEE. The PEE proposal \cite{Wen:2018whg,Wen:2019ubu} may be the most powerful way to calculate the PEE in two dimension theories. Since we heavily rely on this proposal, it is necessary to give a brief review here. The proposal claims that, the PEE is given by a linear combination of certain subset entanglement entropies. This linear combination satisfies the key property of additivity. Furthermore, it was shown to satisfy all the seven requirements using only the general properties of entanglement entropy, thus can be applied to generic theories. Especially in Poincar\'e invariant theories the PEE proposal has been shown to be the unique solution to all the physical requirements \cite{Wen:2019iyq}.

\begin{figure}[h] 
   \centering
   \includegraphics[width=0.5\textwidth]{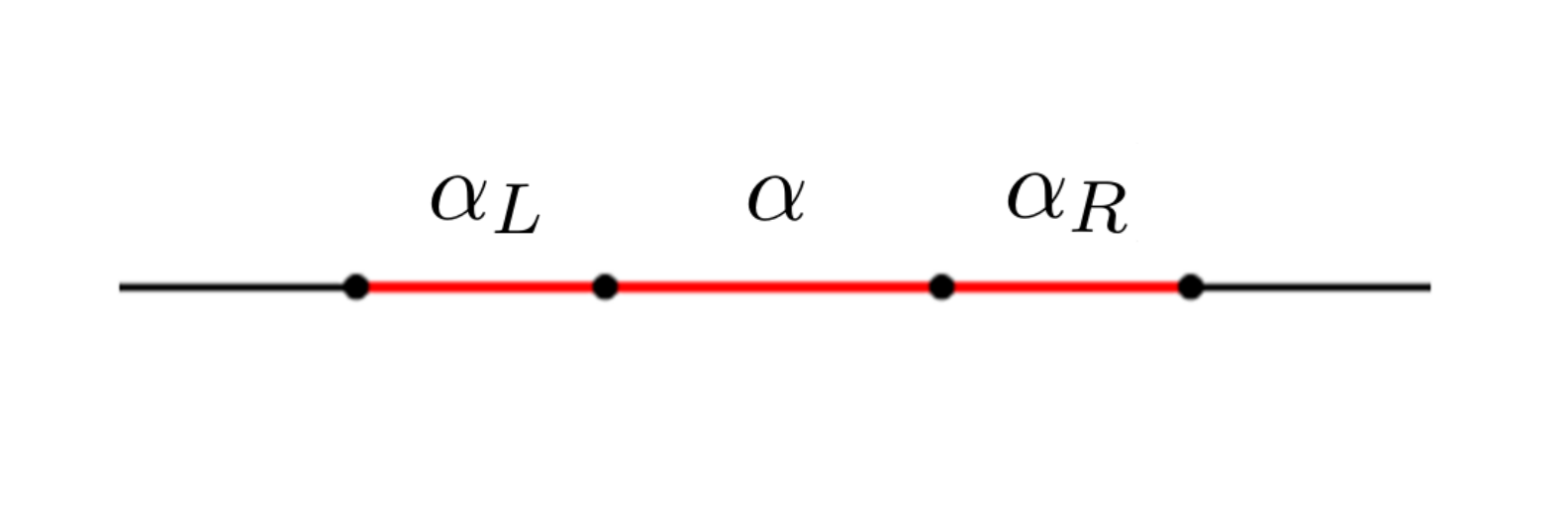} 
 \caption{
\label{order1} The region $A$ is shown by the red interval. When an arbitrary subset $\alpha$ is chosen, a natural decomposition of $A=\alpha_L\cup\alpha\cup\alpha_R$ is determined. All the degrees of freedom in $A$ lines in a definite order.}
\end{figure}

However, a definite order is required for all the degrees of freedom in $A$ for the satisfactory of the additivity. More explicitly, given a region $A$ and an arbitrary subset $\alpha$, when there is a definite order inside $A$, in general it can be partitioned into $A=\alpha_L\cup\alpha\cup\alpha_R$ (see for example Fig.\ref{order1}). Here $\alpha_{L}$ ($\alpha_{R}$) is denoted as the subset on the left (right) hand side of $\alpha$. In this configuration, the PEE proposal claims that
\begin{align}\label{ECproposal}
s_{A}(\alpha)=\frac{1}{2}\left(S_{ \alpha_L\cup\alpha}+S_{\alpha\cup \alpha_R}-S_{ \alpha_L}-S_{\alpha_R}\right)\,.
\end{align}
The order determines $\alpha_L$ and $\alpha_R$ unambiguously. In two-dimensional theories the definite order always exist in the configurations where the region $A$ (whether connected or disconnected) is embedded in a larger one-dimensional chain or circle.

\begin{figure}[h] 
   \centering
   \includegraphics[width=0.5\textwidth]{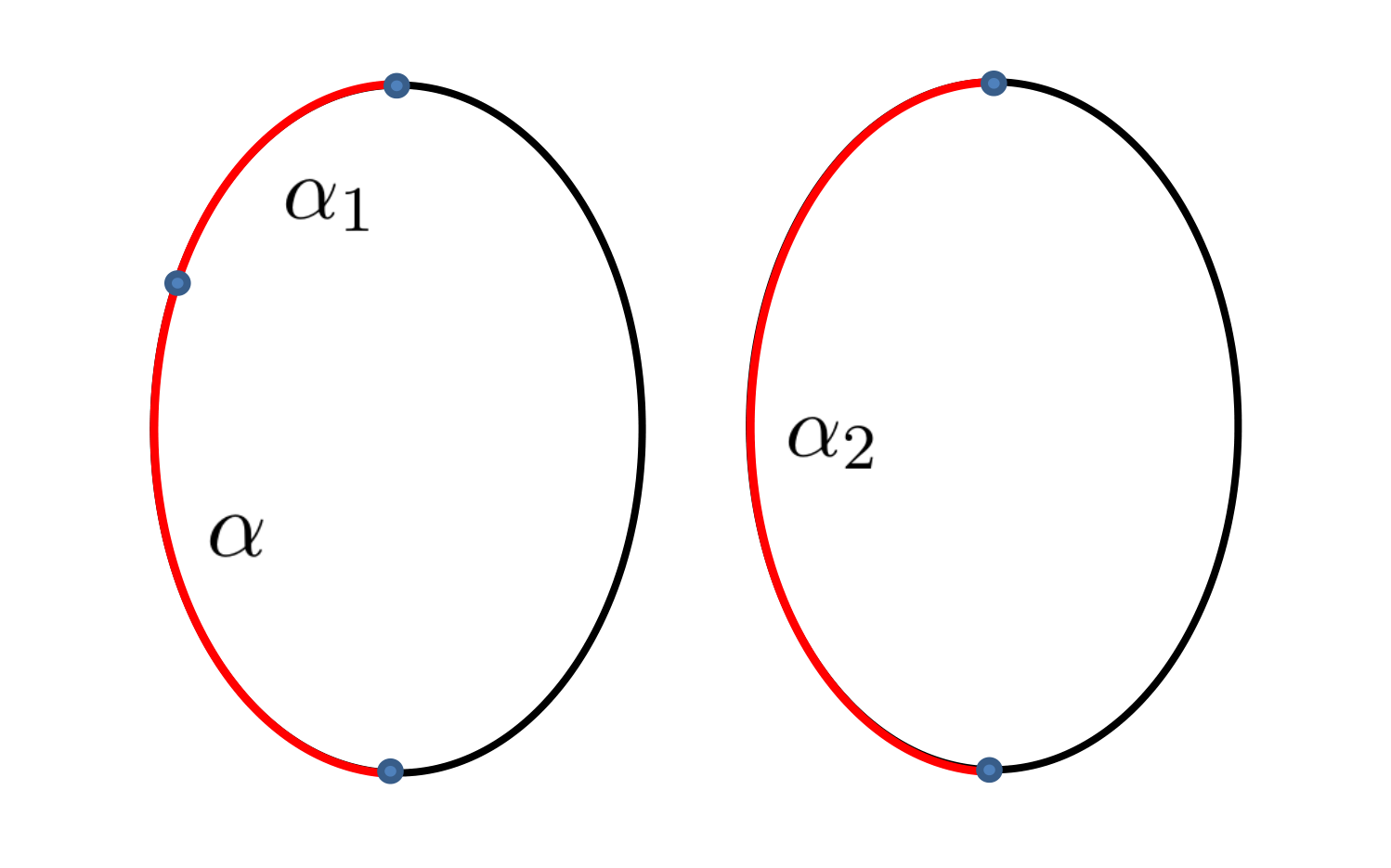}  
 \caption{
\label{noorder1} In this case the region $A$ is the red disconnected multi-interval. Since they lines in different circles, there is no definite order inside $A$. It is not clear whether $\alpha_2$ should be considered as the right-hand side subset of left-hand subset of $\alpha$. }
\end{figure}

Note that, a definite order does not always exist in generic two-dimensional systems. This has not been carefully discussed in previous studies. For example in Fig.\ref{noorder1}, the pure state is settled on two disconnected circles. The region $A$ is given by the two red half circles which are also disconnected. When the subset $\alpha$ is chosen, $A=\alpha\cup\alpha_1\cup\alpha_2$ is partitioned into three parts. In this case, the order between the three parts is ambiguous. One can either take $\alpha_L=\alpha_1\cup\alpha_2\,,\alpha_R=\varnothing$ or $\alpha_L=\alpha_1\,,\alpha_R=\alpha_2$. These two choices represent two different orders. Following \eqref{ECproposal}, the two orders give different values for $s_{A}(\alpha)$, thus the PEE become ambiguous. Also in the case where $A$ is a circle with no boundary, the order is also ambiguous. When the order is not definite, then the proposal \eqref{ECproposal} become ambiguous. In these cases, we should sue to other proposals to calculate the PEE. 

It could be quite useful to make a clarification about the configurations where we can explicitly calculate the PEE or entanglement contour, thus study the BPE.
\begin{enumerate}
\item The entanglement contour for one dimensional regions in general theories with a definite order can be calculated using the PEE proposal \cite{Wen:2018whg,Wen:2019ubu}. The logic of the PEE proposal even works for disconnected intervals with a definite order (see for example \cite{Kudler-Flam:2019nhr}).

\item The entanglement contour for highly symmetric regions in higher dimensions, which can be characterized by a single coordinate, can also be calculated by the PEE proposal \cite{Han:2019scu}. These are called the \textit{quasi-one-dimensional} configurations. For example the contour function for balls and annuli with rotational symmetries, strips with translation symmetries. This also works for general theories.

\item In holographic theories, the entanglement contour for regions with local modular Hamiltonian can be calculated by the geometric construction. This works for the intervals (static or covariant) and balls in higher dimensions \cite{Kudler-Flam:2019oru,Han:2019scu}. It is also valid for holographic theories beyond AdS/CFT (see for example \cite{Wen:2018mev}).

\item In Poincar\'e invariant CFTs with general dimensions, the PEE between any two connected regions and the entanglement contour for any connected region can be calculated by the general formula derived in \cite{Wen:2019iyq}.
\end{enumerate}

In this paper, we mainly use the PEE proposal and the geometric construction to one-dimensional regions. We will also not discuss the Gaussian formula. We will focus on systems in two-dimensional spacetime, especially those with a geometric dual in the context of AdS/CFT. Because in these cases we have more tools to calculate the PEE. The adjacent cases and non-adjacent cases will be discussed separately.

\section{Aspects of BPE when $A$ and $B$ are adjacent}\label{section3}
In the previous section, we defined the \textit{balanced partial entanglement}. Here we claim that the BPE gives the area of the EWCS. In this section we explicitly calculate the BPE in the case of the global AdS$_3$ that duals to the vacuum state of the boundary CFT$_2$. We take $A$ and $B$ to be intervals on the AdS boundary, then the boundary vacuum state is a purification of $\rho_{AB}$. We firstly consider $A$ and $B$ to be adjacent and leave the non-adjacent case for the next section. We will explicitly calculate the BPE$(A,B)$ and compare it with the EWCS. Then we discuss the general entropy relations satisfied by BPE$(A,B)$. Also BPE$(A,B)$ is calculated in the case of the minimal purification in the context of the \textit{\textit{surface/state correspondence}} \cite{Miyaji:2015yva}.

\begin{figure}[h] 
   \centering
   \includegraphics[width=0.4\textwidth]{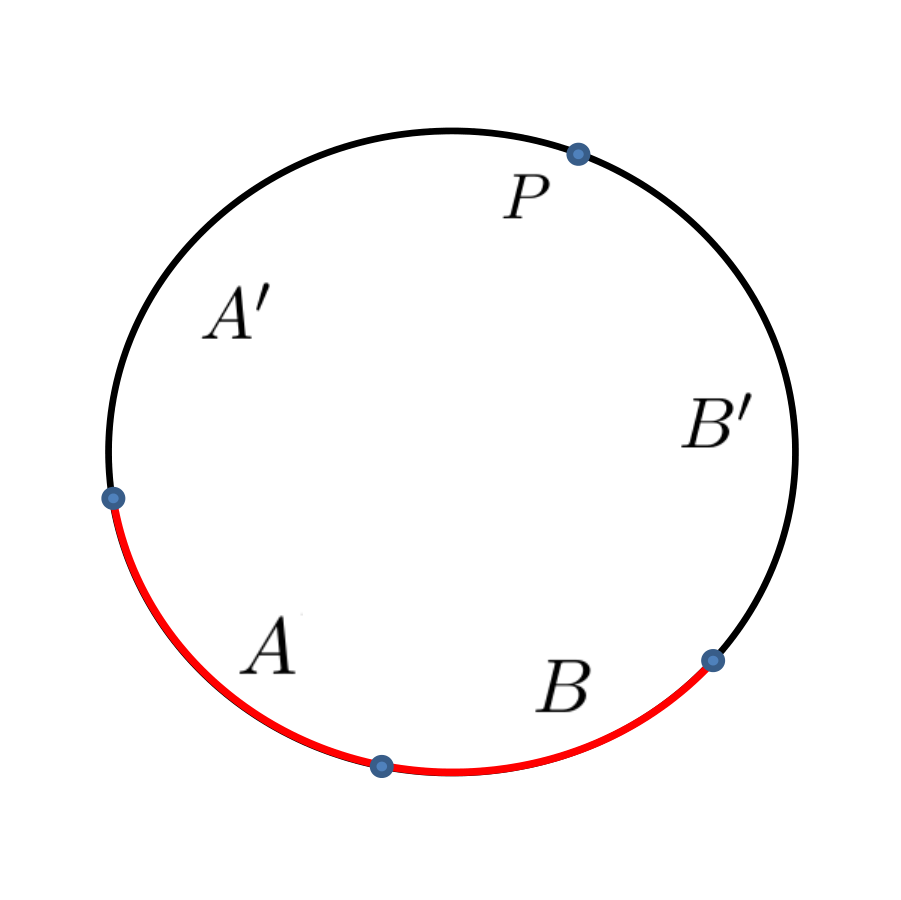} 
 \caption{The above circle is the boundary of the global AdS$_3$ in a time slice, while $AB$ is an interval in the circle. The complement $A'\cup B'$ is partitioned by a point $P$. The balance requirements can be satisfied by adjusting the position of $P$.
\label{adjacentcase} }
\end{figure}

\subsection{Holographic BPE for adjacent intervals}
Let us consider the case in Fig.\ref{adjacentcase}, where the circle is the boundary of the global AdS$_3$. The balance requirement \eqref{requirement1} will determine the position of the point $P$ that partition the complement $A'B'$ into $A'$ and $B'$. Note that, the requirement \eqref{requirement1} does not refer to any information from the bulk geometry.

Here we apply the PEE proposal to calculate the PEE in \eqref{requirement1}. The region here we consider is $AA'$ while the subset is $\alpha=A$. In this case if we define $\alpha_R=A'$, then $\alpha_L=\varnothing$. Following \eqref{ECproposal} we have
\begin{align}\label{PEEAAp}
 s_{AA'}(A)=\frac{1}{2}\left(S_{A}+S_{AA'}-S_{A'}\right)\,.
 \end{align} 
Similarly we have
\begin{align}
s_{BB'}(B)=\frac{1}{2}\left(S_{B}+S_{BB'}-S_{B'}\right)\,.
\end{align}
The balance requirement \eqref{requirement1} then gives the following equation 
\begin{align}\label{BPEa}
S_{A}+S_{AA'}-S_{A'}=&S_{B}+S_{BB'}-S_{B'}\,,
\cr
\Rightarrow~~~~~~~~~~~~~~~~
S_A-S_B=&S_{A'}-S_{B'}\,,
\end{align}
 which is enough to determine the point $P$. 
 
In global AdS$_3$, the entanglement entropy for an arbitrary interval of length $l$ is given by 
\begin{align}\label{EEinterval}
S=\frac{c}{3}\log\left(\frac{2}{\epsilon }\sin\frac{ l}{2}\right)\,,
\end{align}
where we have taken the length of the AdS boundary circle to be $2\pi$, the AdS radius $\ell=1$ and $c=\frac{3}{2G}$. Let us use $l_A,\, l_B,\, l_{A'}$, and $l_{B'}$ to denote the length of all the relevant intervals. Obviously, we have
\begin{align}
 l_A+ l_B +l_{A'}+l_{B'}=&2\pi\,,
 \end{align} 
also \eqref{BPEa} can be written as
\begin{align}
\frac{\sin( l_A/2)}{\sin( l_B /2)}=&\frac{\sin( l_{A'}/2)}{\sin(  l_{B'}/2)}\,.
\end{align}
The solution of the above two equations is given by
\begin{align}\label{solutionlAp}
l_{A'}=&-2 \cot ^{-1}\left[\frac{1}{2} \csc \left(\frac{{l_A}}{2}\right) \left(\sin \left({l_A}+\frac{{l_B}}{2}\right)-3 \sin \left(\frac{{l_B}}{2}\right)\right) \csc \left(\frac{{l_A}}{2}+\frac{{l_B}}{2}\right)\right]\,,
\cr
l_{B'}=&~2\pi-l_A-l_B-l_{A'}\,.
\end{align}
Plugging the solution into the PEE \eqref{PEEAAp}, we get the BPE,
\begin{align}\label{adjacentepab}
\text{BPE}(A,B)=\frac{c}{6}\log \left[\frac{4}{\epsilon}\frac{\sin(l_A/2)\sin(l_B/2)}{\sin((l_A+l_B)/2)}\right]\,.
\end{align}
One can check that, the $\text{BPE}(A,B)$ exactly gives the area of the EWCS calculated in \cite{HEoP1,HEoP2}.

The above calculation shows that the BPE captures the specific correlation between $A$ and $B$, which is represented by the area of the EWCS. Since the BPE can be defined in general quantum system, the BPE could be considered as a generalization of the EWCS to non-holographic systems. 

More interestingly we find that the crossing PEE $\mathcal{I}_{AB'}$ in this case is a constant independent from $l_A$ and $l_B$. Using the PEE proposal and the solution \eqref{solutionlAp}, we find
\begin{align}
\mathcal{I}_{AB'}=s_{A'AB}(A)=&\frac{1}{2}\left(S_{A'A}+S_{AB}-S_{A'}-S_{B}\right)
\cr
=&\frac{c}{6}\log \left[\frac{\sin\left((l_A+l_{A'})/2\right)\sin\left((l_A+l_{B})/2\right)}{\sin\left(l_{A'}/2\right)\sin\left(l_{B}/2\right)}\right]
\cr
=&\frac{c}{6}\log 2\,.
\end{align}
This is surprising that the crossing PEE is independent from the partition of the pure state, hence can be used to characterize or classify the purifications. Later we will show that the crossing PEE for the canonical purification is also $\frac{c}{6}\log 2$. However, there is no evidence that the crossing PEE is invariant under all the purifications.

\subsection{Entropy relations for BPE}
The PEEs which can be written as a linear combination of the entanglement entropies $S_A,~S_B$ and $S_{AB}$ are of course purification independent. For example,
\begin{align}\label{peemi}
\mathcal{I}(A,B)&=\frac{1}{2}I(A,B)\,,
\\
\mathcal{I}(A,A'B')&=\frac{1}{2}\left(S_A+S_{AB}-S_{B}\right)\,,
\\
\mathcal{I}(B,A'B')&=\frac{1}{2}\left(S_B+S_{AB}-S_{A}\right)\,.
\end{align}
Note that the relation \eqref{peemi} between $\mathcal{I}(A,B)$ and the mutual information $I(A,B)$ only holds for the adjacent cases. The balance requirements give one more purification independent quantity 
\begin{align}
\left(\mathcal{I}_{AA'}-\mathcal{I}_{BB'}\right)=S_A-S_B.
\end{align} 
Unfortunately the crossing PEE, as well as BPE$(A,B)$, is not purification independent. BPE$(A,B)$ is only invariant under the unitary transformations which keep $\mathcal{I}_{AB'}$ fixed. These include
\begin{itemize}
\item the local unitary transformations on $A'$ and $B'$ respectively,

\item the unitary transformations that adding or removing correlations between $A'$ and $B'$,

\item the unitary transformations that adding or removing correlation between $A$ and $A'$ while adding or removing the same amount of correlation between $B$ and $B'$, i.e. keeping $\mathcal{I}_{AA'}-\mathcal{I}_{BB'}$ fixed.
\end{itemize}
Note that, the correlation here means the PEE.

Now we discuss the general entropy relations satisfied by BPE$(A,B)$. The upper bound of the PEE indicates $s_{AA'}(A)\leq S_{A}$ and $s_{BB'}(B)\leq S_{B}$. Imposing the balance requirement, we directly get
\begin{align}\label{property1}
property~1:~\text{BPE}(A,B)\leq \text{min}(S_A,S_B)\,.
\end{align}
The above relation can be satisfied in general cases.

When the balance requirements are satisfied, BPE$(A,B)=\mathcal{I}_{AB}+\mathcal{I}_{AB'}$. In the adjacent cases we also have \eqref{peemi}. Then the positivity of the PEE directly gives the following relation
\begin{align}\label{property2}
property~2:~\text{BPE}(A,B)\geq \frac{1}{2}I(A,B).
\end{align}
Note that in the non-adjacent cases the above relation cannot be proved in the same way. In terms of the PEEs, the entanglement entropy satisfies the following decomposition
\begin{align}
S_A=\mathcal{I}_{AB}+\mathcal{I}_{AA'}+\mathcal{I}_{AB'}\,,
\qquad
S_B=\mathcal{I}_{AB}+\mathcal{I}_{BA'}+\mathcal{I}_{BB'}\,.
\end{align}
In the cases where $AB$ is a connected region, we also have
\begin{align}\label{sabpee}
S_{AB}=&\mathcal{I}_{AA'}+\mathcal{I}_{AB'}+\mathcal{I}_{BB'}+\mathcal{I}_{BA'}\,.
\end{align}
One can easily see that the above decompositions directly gives
\begin{align}
I(A,B)=S_A+S_B-S_{AB}=2\mathcal{I}(A,B).
\end{align}
However the decomposition \eqref{sabpee} is not accurate for the non-adjacent cases\footnote{Note that this evaluation of entanglement entropy using PEE is very subtle for disconnected regions. For example, previous studies \cite{Berthiere:2019lks,Roy:2019gbi,Wen:2019iyq} showed that naively taking an uniform cutoff for all the endpoints of multi-intervals in CFT$_2$ will give the results of Refs. \cite{Casini:2005rm,Casini:2004bw,Calabrese:2004eu,Hubeny:2007re}, which is only justified for free fermions while incorrect in more general theories. So the relation $I(A,B)=2\mathcal{I}(A,B)$ only holds when $AB$ is a connected region.}, thus the comparison between $\text{BPE}(A,B)$ and $\frac{1}{2}I(A,B)$ is not clear.

The properties \eqref{property1} and \eqref{property2} directly lead to other interesting entropy relations. For example the polygamy inequality for a the system $ABC$ in a pure state,
\begin{align}
property~3:~\text{BPE}(A,B)+\text{BPE}(A,C)\geq \text{BPE}(A,BC)\,.
\end{align}
and the saturation of the upper bound when the Araki-Lieb inequality is saturated
\begin{align}
property~4:~|S_A-S_B|=S_{AB}~~\Rightarrow~~\text{BPE}(A,B)=\min(S_A,S_B)\,.
\end{align}

The monotonicity of the BPE$(A,B)$ can also be justified using the additivity and positivity of the PEE. We assume that the partition point $P$ satisfies the balance requirement.  Let us consider a region $C$ which is inside $B'$ and adjacent to $B$. We define the complement of $C$ inside $B'$ to be $B''$ thus $B'=C\cup B''$. According to the additivity and positivity we have 
\begin{align}
s_{AA'}(A)=s_{BB'}(B)<s_{BB'}(BC)\,.
\end{align}
Then we combine $B$ and $C$, and consider BPE$(A,BC)$. Since $B$ expands to $BC$, the balance requirement is now $s_{AA'}(A)=s_{BB'}(BC)$. We need to adjust the position of $P$ to go back to balance. In other words we should increase $s_{AA'}(A)$ and reduce $s_{BB'}(BC)$ by adjusting $P$. Since PEE is additive, we can write
\begin{align}
s_{AA'}(A)&=\mathcal{I}(A,BC)+\mathcal{I}(A,B'')
\cr
s_{BB'}(BC)&=\mathcal{I}(BC,A)+\mathcal{I}(BC,A')
\end{align}
Our goal can be easily achieved by moving $P$ towards $A$, thus $B''$ expands while $A'$ shrinks. Due to the positivity and additivity, this procedure increases $\mathcal{I}(A,B'')$ and reduces $\mathcal{I}(BC,A')$. After $P$ is properly settled down such that $s_{AA'}(A)=s_{BB'}(BC)$, the PEE $s_{AA'}(A)=\text{BPE}(A,BC)$ is bigger than its previous value $\text{BPE}(A,B)$. In other words, we get the monotonicity of BPE,
\begin{align}
property~5:~\text{BPE}(A,BC)\geq\text{BPE}(A,B)\,.
\end{align}
In the cases where $A$ and $B$ are non-adjacent, the partition point is more than one. A similar argument also lead to the monotonicity using the positivity and additivity of the PEE.

In summary our arguments for properties 1 and 5 also applies for non-adjacent cases. The argument for property 2 applies to the adjacent cases in a generic theory. The properties 3 and 4 follows from properties 1 and 2. The properties 1-5 are all satisfied by the EoP.

\subsection{BPE in the minimal purification and the entanglement of purification}
Then we consider another purification, which is closely related to the EoP. The EoP is defined in the following: assuming a bipartite system $A B$ is in a mixed state $\rho_{AB}$. Let $\ket{\psi}\in \mathcal{H}_{AA'}\otimes \mathcal{H}_{BB'}$ be a purification of $\rho_{AB}$. The EoP $E_{p}(A,B)$ \cite{EoP} is defined by:
\begin{align}
E_{p}(A,B)=\mathop{\text{min}}\limits_{\phi,A'}S_{AA'}\,,
\end{align}
where minimization is over all the purifications and all the partitions of $A'B'$. The EoP is an intrinsic entanglement measure independent from purifications.

It was proposed \cite{HEoP1,HEoP2} that in the context of AdS/CFT the EoP $E_{p}(A,B)$ is dual to the area of the EWCS $\Sigma_{AB}$ in the following way
\begin{align}
 E_{p}(A,B)=\frac{Area(\Sigma_{AB})}{4G}
 \end{align} 
Also the entropy relations satisfied by $\Sigma_{AB}$ are the same with those satisfied by EoP. Though it is hard to prove this duality, in the context of the \textit{\textit{surface/state correspondence}} \cite{Miyaji:2015yva} the calculation of the area of $\Sigma_{AB}$ perfectly agrees with the definition of the EoP \cite{HEoP1,Guo:2019azy}. The \textit{\textit{surface/state correspondence}} proposes much more general correspondence between bulk codimension-2 convex \cite{Miyaji:2015yva} spacelike surfaces and quantum states, based on the tensor network description of the AdS/CFT correspondence. In the case of AdS$_3$/CFT$_2$ the main points of the \textit{\textit{surface/state correspondence}} are in the following.
\begin{itemize}

\item Convex curves that homologous to a point correspond to pure states (for example circles in the bulk with no black hole inside), otherwise the curves correspond to mixed states (like intervals in the bulk or circles that surrounding a black hole).

\item Any two surfaces $\sigma_1$ and $\sigma_2$ that are connected by a smooth deformation preserving convexity, are related by an unitary transformation.

\item The entanglement entropy for any convex curves $\sigma$ are calculated by the area of the minimal surfaces that homologous $\sigma$, which is a straight forward generalization of the RT formula.
\end{itemize}

\begin{figure}[h] 
   \centering
   \includegraphics[width=0.45\textwidth]{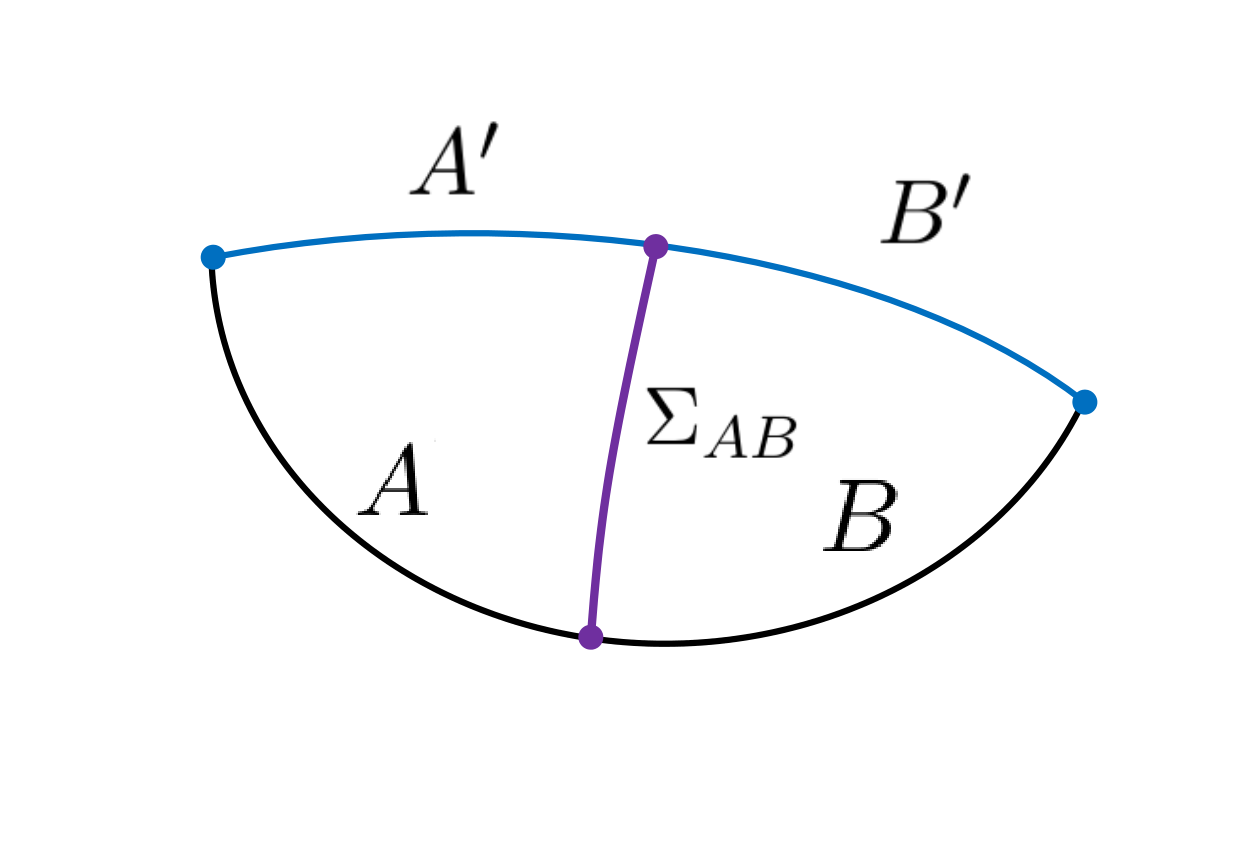} 
 \caption{Here $A'B'$ is deformed to be the RT surface of $AB$, it is also the RT surface of itself. $S_{AA'}$ reaches its minimal value when $\mathcal{E}_{AA'}$ coincide with $\Sigma_{AB}$, which is the purple line.
\label{surfacestatep} }
\end{figure}

In the case of global AdS$_3$, the boundary $ABA'B'$ is a circle in a pure state. Let us consider the simple case of Fig.\ref{adjacentcase}. In the context of the \textit{\textit{surface/state correspondence}} it is convenient to perform unitary transformations on $A'B'$ by deforming of the curve $A'B'$ while keeping the endpoints fixed. Among all the deformations and partitions, $S_{AA'}$ arrives at its minimal value when $A'B'$ is deformed to the RT surface $\mathcal{E}_{AB}$ and $A'B'$ is properly partitioned thus $\mathcal{E}_{AA'}$ is normal to $\mathcal{E}_{A'B'}=A'B'$. It is easy to find that $\mathcal{E}_{AA'}$ coincides with $\Sigma_{AB}$, (see Fig.\ref{surfacestatep}). Hence the calculation of the length of $\Sigma_{AB}$ can be achieved via the minimization of $S_{AA'}$ among all the purifications and partitions. 

Then let us consider the configuration of Fig.\ref{surfacestatep} without specifying the partition of $A'B'$. We can evaluate the BPE$(A,B)$ using the PEE proposal and balance requirements. Since $A'B'$ is a minimal surface, using the generalized RT formula for the bulk convex curves we have
\begin{align}
S_{A'}=\frac{l_{A'}}{4G}\,,\qquad S_{B'}=\frac{l_{B'}}{4G}\,.
\end{align}
Since $\mathcal{E}_{AB}=A'\cup B'$ we have
\begin{align}
S_{A'}+S_{B'}=S_{AB}\,.
\end{align}
The balance requirement requires that
\begin{align}
S_{A'}-S_{B'}=S_A-S_B
\end{align}
Solving the above two equations we have
\begin{align}\label{solutionssc}
S_{A'}=\frac{c}{6}\log\left[\frac{2\sin\frac{l_A}{2}\sin\frac{l_A+l_B}{2}}{\epsilon\sin\frac{l_B}{2}}\right]\,,
\qquad 
S_{B'}=\frac{c}{6}\log\left[\frac{2\sin\frac{l_B}{2}\sin\frac{l_A+l_B}{2}}{\epsilon\sin\frac{l_A}{2}}\right]\,.
\end{align}
The above solution determines the position of the partition point $P$. It is interesting that the point $P$ is exactly where $\Sigma_{AB}$ intersecting with $\mathcal{E}_{AB}$. One may wonder that here the BPE$(A,B)$ may also give the length of $\Sigma_{AB}$. This is obviously not true because
\begin{align}
 S_{AA'}>s_{AA'}(A)\,.
 \end{align} 
 Also according to the generalized RT formula and \eqref{adjacentepab} we have
 \begin{align}
S_{AA'}=\frac{Area(\Sigma_{AB})}{4G}=\frac{c}{6}\log \left[\frac{4}{\epsilon}\frac{\sin(l_A/2)\sin(l_B/2)}{\sin((l_A+l_B)/2)}\right]\,,
 \end{align}
while the BPE$(A,B)$ is given by 
\begin{align}
\text{BPE}(A,B)=&s_{AA'}(A)=\frac{1}{2}\left(S_{A}+S_{AA'}-S_{A'}\right)
\cr
=&\frac{c}{6}\log\left[\frac{2\sin\frac{l_A}{2}\sin\frac{l_B}{2}}{\epsilon\sin\frac{l_A+l_B}{2}}\right]+\frac{c}{12}\log 2
\cr
=&\mathcal{I}_{AB}+\frac{c}{12}\log 2\,.
\end{align}
Here we used the solution \eqref{solutionssc}. Also one can easily read that the crossing PEE is just given by
\begin{align}
 \mathcal{I}_{AB'}=\frac{c}{12}\log 2\,,
 \end{align} 
which is again a constant independent from $l_A$ and $l_B$. However, it is different from the constant $\frac{c}{6}\log 2$ for the case of the global AdS$_3$. In other words, in the context of the \textit{\textit{surface/state correspondence}} the unitary transformation, that evolve the boundary vacuum state of the global AdS$_3$ to the ``minimal'' purification shown in Fig.\ref{surfacestatep}, changes the crossing PEE $\mathcal{I}_{AB'}$ hence changes the BPE$(A,B)$.

\section{BPE for non-adjacent intervals}\label{section4}
 
\begin{figure}[h] 
   \centering
   \includegraphics[width=0.45\textwidth]{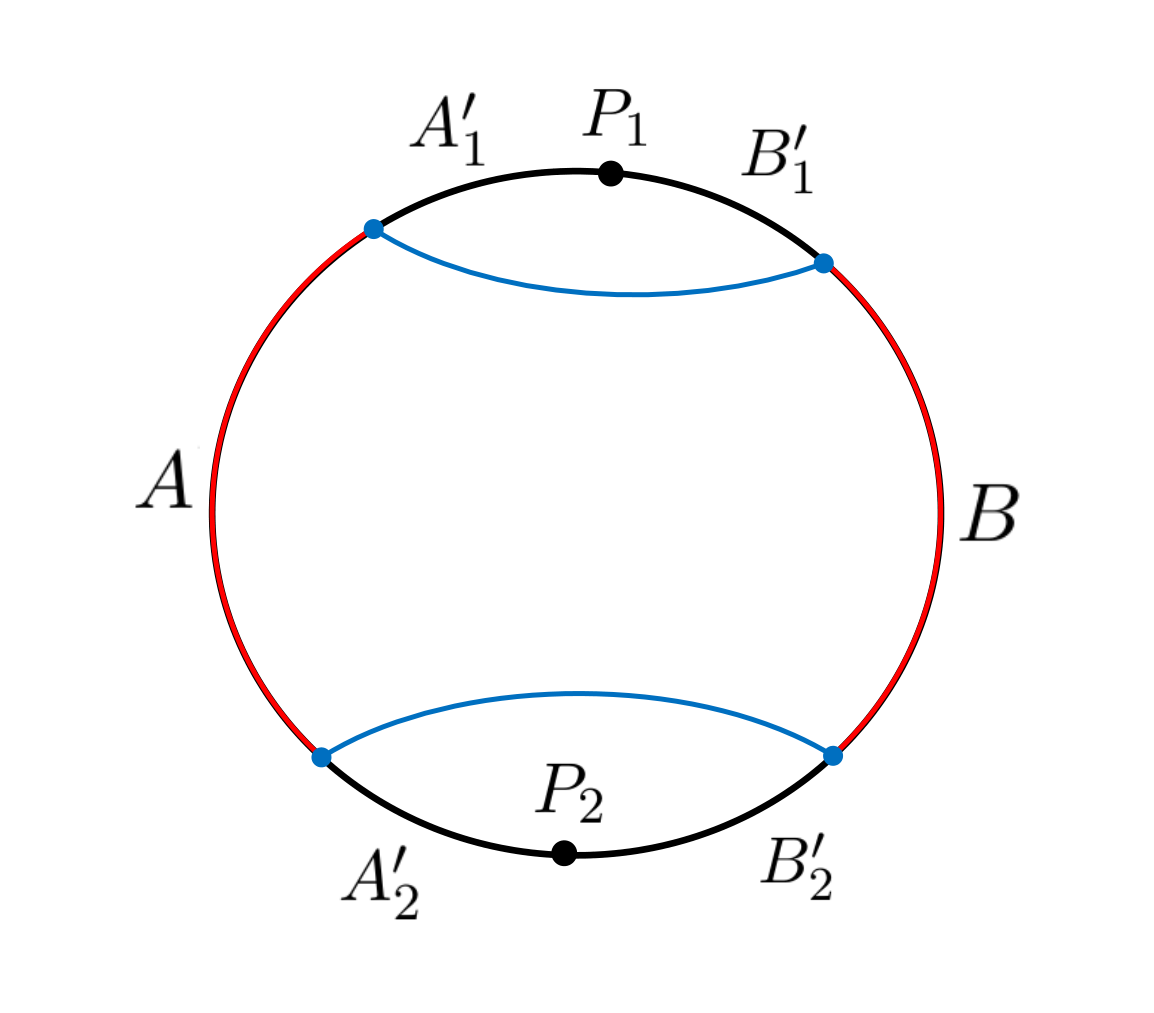} 
 \caption{The region $AB$ is disconnected two-interval with connected entanglement wedge. The partition of the combined system $ABA'B'$ follows our prescription to minimize the $s_{AA'}(A)$ and is determined by the position of two points $P_1$ and $P_2$ under the balance requirements,.
\label{nonadjacent} }
\end{figure}

Then we consider the cases where $A$ and $B$ are non-adjacent intervals on the AdS boundary and the entanglement wedge $\mathcal{W}_{AB}$ is connected. For example, see Fig.\ref{nonadjacent}. In this case the compliment $A'B'$ of $AB$ on the boundary is also disconnected. It is partitioned by two points $P_1$ and $P_2$ into four regions that are classified into two pairs. Let us denote, for example, the region partitioned by $P_1$ as $A'_1\cup B'_1$. As we have mentioned previously, the position of $P_1$ and $P_2$ can be determined by the balance requirements \eqref{BR2}. Here we rewrite the two independent requirements in the following
\begin{align}\label{balancer2}
s_{AA'}(A'_1)=s_{BB'}(B'_1)\,, \qquad s_{AA'}(A'_2)=s_{BB'}(B'_2)\,.
\end{align} 
Note that $s_{AA'}(A)=s_{BB'}(B)$ follows from $S_{AA'}=S_{BB'}$. 

Let us denote the length of the intervals to be
\begin{align}
l_A=2a,\quad l_B=2b, \quad l_{A_1'}=2a_1,\quad l_{B_1'}=2b_1\quad  l_{A_2'}=2a_2,\quad l_{B_2'}=2b_2.
\end{align}
Given the length and position of $A$ and $B$, the length and position of $A_1'B_1'$ and $A_2'B_2'$ are also determined. Assuming $a_1+b_1=\alpha$ we have $a_2+b_2=2\pi-2a-2b-2\alpha$. Since $\alpha$ is known to us, there remains only two undetermined parameters. Let us take them to be $a_1$ and $a_2$ thus
\begin{align}\label{b1b2}
 b_1=\alpha-a_1\,, \qquad b_2=\pi-\alpha-a-b-a_2\,.
\end{align} 
The balance requirements \eqref{balancer2} give that
\begin{align}
S_{A_1'}-S_{B_1'}=S_{AA_2'}-S_{BB_2'}\,,
\qquad
S_{A_2'}-S_{B_2'}=S_{AA_1'}-S_{BB_1'}\,.
\end{align}
Using the holographic result for entanglement entropy \eqref{EEinterval} and the relations \eqref{b1b2}, the above equations can be rewritten as,
\begin{align}\label{BPEC1}
\frac{\sin[a_1]}{\sin[\alpha-a_1]}=&\frac{\sin[a+a_2]}{\sin[\alpha+a+a_2]}\,,
\\\label{BPEC2}
\frac{\sin[a_2]}{\sin[\alpha+a+b+a_2]}=&\frac{\sin[a+a_1]}{\sin[a+a_2]}\,,
\end{align}
which uniquely determine the partition points of $A'B'$. The solutions are in the following,
\begin{align}\label{solutiona1}
a_1= \cos ^{-1}&\left[\frac{\sin (a-\alpha +a_2)+3 \sin (a+\alpha +a_2)}{\sqrt{2} \sqrt{-2 \cos (2 (a+\alpha +a_2))-2 \cos (2 (a+a_2))+\cos (2 \alpha )+3}}\right]\,,
\\\label{solutiona2}
a+2 a_2= \tan ^{-1}&\Big[\sin (a-b) (\sin (a) \cos (\eta )-\sin (b))-2 \lambda  \sin (a) \sin (\eta ),
\cr
&~~~~~~~~~~-\sin (a) \sin (\eta ) \sin (a-b)-2 \lambda  \sin (a) \cos (\eta )+2 \lambda  \sin (b)\Big]\,,
\end{align}
where
\begin{align}
\lambda=&\sqrt{\sin (a) \sin (b) \sin (a+\alpha ) \sin (\alpha +b)}\,,
\cr
\eta=&a+b+2\alpha\,.
\end{align}
 
Following the PEE proposal we have
\begin{align}
s_{AA'}(A)&=\frac{1}{2}\left(S_{AA_1'}+S_{AA_2'}-S_{A_1'}-S_{A_2'}\right)
\cr
&=\frac{c}{6}\log\left[\frac{\sin[a+a_1]\sin[a+a_2]}{\sin[a_1]\sin[a_2]}\right]\,.
\end{align}
The above PEE gives the BPE$(A,B)$ when we plug in the solutions \eqref{solutiona1}-\eqref{solutiona2}. At last, we find
\begin{align}
\text{BPE}(A,B)&=\frac{c}{6}  \log \left[2  \frac{\lambda+\sin (a) \sin (b)}{\sin (\alpha )\sin (a+\alpha +b)} +1\right]
\end{align}
Though the solution to the balance requirements is a bit complicated, the BPE has a simple expression. One can check that it exactly matches with the length of $\Sigma_{AB}$ previously calculated in \cite{HEoP2}. Compare with the calculation of \cite{HEoP2}, our requirements are simple and donot have to refer to any information from the bulk geometry. Later we will explain why this matching appears using the holographic picture for entanglement contour.

The BPE also satisfies certain entropy relations when $A$ and $B$ are non-adjacent. The property 1 holds in general. The property 2 is not easy to prove for disconnected $AB$ in a generic purification. This may due to the disadvantage that our understanding of the entanglement contour for disconnected regions is not clear \cite{Wen:2019iyq}. However, for holographic cases, because the mutual information is monogamous \cite{Hayden:2011ag}, it was proved in \cite{Wen:2019ubu} that the PEE satisfies the following inequality,
\begin{align}
s_{AA'}(A)\geq \frac{1}{2}I(A,BB')\geq \frac{1}{2}I(A,B)\,.
\end{align}
Since the BPE is also a PEE we have $\text{BPE}(A,B)\geq \frac{1}{2}I(A,B)$. 

For non-adjacent cases, so far we donot have the proof for property 2 for non-holographic theories. We want to point out that, the monogamy of the mutual information is not a necessary condition for the property 2. So it is still possible to prove it in general cases. We leave this point for future study. Since the property 3 and 4 follow from property 2, they are also only justified for holographic theories. While the property 5 of monotonicity can be understood for generic configurations using the similar arguments in the previous section.

\section{The canonical purification and the reflected entropy}\label{section5}
The canonical purification discussed in \cite{Dutta:2019gen} is another example where we can explicitly study the BPE. In \cite{Dutta:2019gen} a new quantity named the reflected entropy was defined and its holographic relation to the EWCS was established. In this section, we will show that the reflected entropy is indeed identical to the BPE for canonical purifications. Consequently the relation between the BPE and $\Sigma_{AB}$ in the canonical purification follows directly. Note that the reflected entropy is only defined in the canonical purification cases, while the BPE can be defined for a generic purification. The BPE can be considered as a generalization of the reflected entropy for generic purifications.

Let us firstly give a brief review on the canonical purification and the reflected entropy. Consider a bipartite system $A\cup B$ with the Hilbert space $\mathcal{H}_{AB}$ and the orthonormal bases $\{\ket{\psi_{i}}\}$. The system is in a mixed state 
\begin{align}
\rho_{AB}=\sum\limits_{i}p_i \ket{\psi_{i}}\bra{\psi_{i}}\,.
\end{align}
Then we introduce a system $A'\cup B'$ with the same copy of the Hilbert space, and the partition is just a reflection of the partition of $AB$. The canonical purification is given by the following pure state for the combined system $ABA'B'$,
\begin{align}
\ket{\sqrt{\rho_{AB}}}=\sum\limits_{i}\sqrt{p_i} \ket{\psi_{i}}_{AB}\ket{\psi_{i}}^*_{A'B'}\,,
\end{align}
where $\{\ket{\psi_{i}}^*\}$ is another orthonormal basis of $\mathcal{H}_{AB}$. Now the mixed state is the reduced density matrix $\rho_{AB}=\text{Tr}_{A'B'}\ket{\sqrt{\rho_{AB}}}\bra{\sqrt{\rho_{AB}}}$. The thermo-field double state is a simple case of the canonical purification. The reflected entropy is then defined as the von Neumann entropy (or entanglement entropy) for $AA'$,
\begin{align}
S_{R}(A:B)=S_{AA'}=-\text{Tr}\rho_{AA'}\log \rho_{AA'}\,,
\end{align}
where $\rho_{AA'}=\text{Tr}_{BB'}\ket{\sqrt{\rho_{AB}}}\bra{\sqrt{\rho_{AB}}}$.

\begin{figure}[h] 
   \centering
         \includegraphics[width=0.45\textwidth]{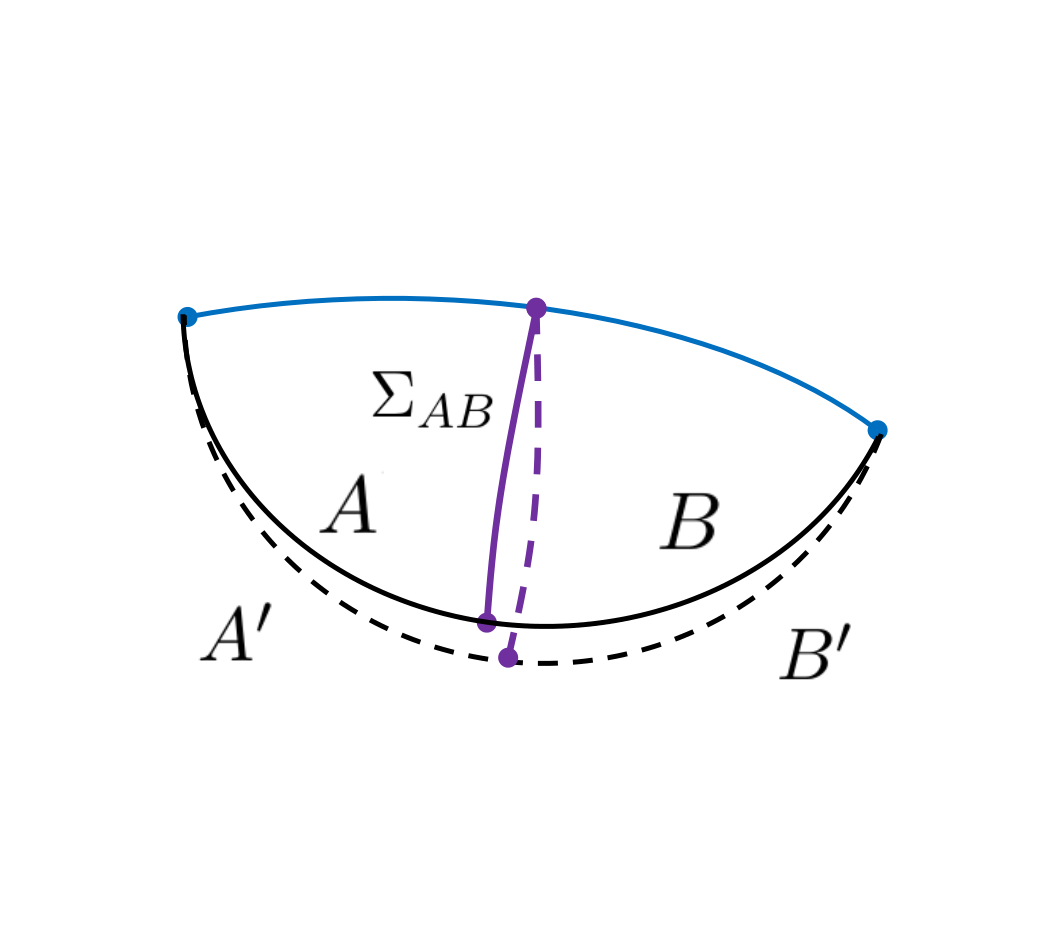}

 \caption{The homology surfaces $\mathcal{R}_{AB}$ and $\mathcal{R}_{A'B'}$ are glued at the RT surface $\mathcal{E}_{AB}$, which is the blue line. This configuration $\mathcal{RR'}_{AB}$ is the gravity dual of the canonical purification of $\rho_{AB}$ and has the reflection symmetry at $\mathcal{E}_{AB}$. The solid purple line is the $\Sigma_{AB}$ while the dashed purple line is its image under reflection.
\label{peeandeop4} }
\end{figure}

It is important that, for holographic systems the canonical purification has a bulk geometric description. For example the eternal black hole \cite{Maldacena:2001kr} describes the thermo-field double state. For more generic cases where $AB$ is a subsystem of a holographic boundary state, the geometric dual for the canonical purification is the manifold glued from the homology surface $\mathcal{R}_{AB}$ and its CPT conjugate $\mathcal{R}_{A'B'}$ along the RT surfaces $\mathcal{E}_{AB}$ and $\mathcal{E}_{A'B'}$. For example see Fig.\ref{peeandeop4}. This glued manifold, denoted as $\mathcal{RR'}_{AB}$, is the gravity dual of the canonical purification $\ket{\rho_{AB}}$. By construction it has reflection symmetry at the RT surface $\mathcal{E}_{AB}$. This holographic construction is proposed in \cite{Dutta:2019gen} using the Engelhardt-Wall procedure \cite{Engelhardt:2018kcs,Engelhardt:2017aux}. The entanglement entropy $S_{AA'}$ is then holographically calculated by the minimal surface $\mathcal{E}_{AA'}$ in $\mathcal{RR'}_{AB}$ that anchored on the boundary of $AA'$. 

It was shown in \cite{Dutta:2019gen} that $\mathcal{E}_{AA'}$ is closely related to $\Sigma_{AB}$. In general $\Sigma_{AB}$ can be determined by the following two properties. Firstly it should be a geodesic chord in $\mathcal{R}_{AB}$ that separates $A$ from $B$. Secondly it should be anchored on the RT surface $\mathcal{E}_{AB}$ vertically, which is required by the reflection symmetry. For example, in Fig.\ref{peeandeop4} where $A$ and $B$ are adjacent, $\Sigma_{AB}$ is the geodesic chord that emanates from the joint point of $A$ and $B$ and end on $\mathcal{E}_{AB}$ vertically, which is shown by the solid purple line. The dashed purple line is the image of $\Sigma_{AB}$ under reflection. The RT surface $\mathcal{E}_{AA'}$ is just formed by $\Sigma_{AB}$ and its image. This directly gives that
\begin{align}\label{REMSC}
 Area(\mathcal{E}_{AA'})=2Area(\Sigma_{AB})\,.
 \end{align} 
 The reflected entropy is then related to $\Sigma_{AB}$ in the following way
 \begin{align}
 \frac{1}{2}S_{R}(A,B)=\frac{ Area(\mathcal{E}_{AA'})}{8G}=\frac{ Area(\Sigma_{AB})}{4G}
 \end{align}

Now let us calculate the BPE$(A,B)$. We see that in Fig.\ref{peeandeop4} the combined system $ABA'B'$ forms a circle and $AA'$ is a connected interval with a definite order. In this case we can calculate the PEE $s_{AA'}(A)$ using the PEE proposal \eqref{ECproposal},
\begin{align}
s_{AA'}(A)=\frac{1}{2}(S_{A}+S_{AA'}-S_{A'})=\frac{1}{2}S_{AA'}.
\end{align}
We used $S_{A}=S_{A'}$ in the above equation, which follows from the reflection symmetry. However the definite order for $AA'$ is not guaranteed for generic configurations. Fortunately there exists a generic way to calculate $s_{AA'}(A)$ in canonical purification cases using the reflection symmetry. The reflection symmetry indicates that the contribution from $A$ and $A'$ to $S_{AA'}$ are equal. Regarding the normalization property that $s_{AA'}(A)+s_{AA'}(A')=S_{AA'}$, in general we have
\begin{align}\label{saapcanonical}
s_{AA'}(A)=s_{AA'}(A')=\frac{1}{2}S_{AA'}=\frac{1}{2}S_{R}(A,B).
\end{align}
Similarly we have
\begin{align}
s_{BB'}(B)=s_{BB'}(B')=\frac{1}{2}S_{BB'}.
\end{align}
Since $S_{AA'}=S_{BB'}$, we straightforwardly find
\begin{align}
s_{AA'}(A)=s_{BB'}(B)\,,
\end{align}
which is exactly the balance requirement. In summary in the canonical purifications for any $\rho_{AB}$ where the partition of $A'B'$ is a reflection of the partition of $AB$, we have $s_{AA'}(A)=\text{BPE}(A,B)$.  This further more indicates that the BPE$(A,B)$ is directly related to the EWCS,
\begin{align}\label{epabsr}
\text{BPE}(A,B)=\frac{1}{2}S_{R}(A,B)=\frac{ Area(\Sigma_{AB})}{4G}\,.
\end{align}

\begin{figure}[h] 
   \centering
\includegraphics[width=0.45\textwidth]{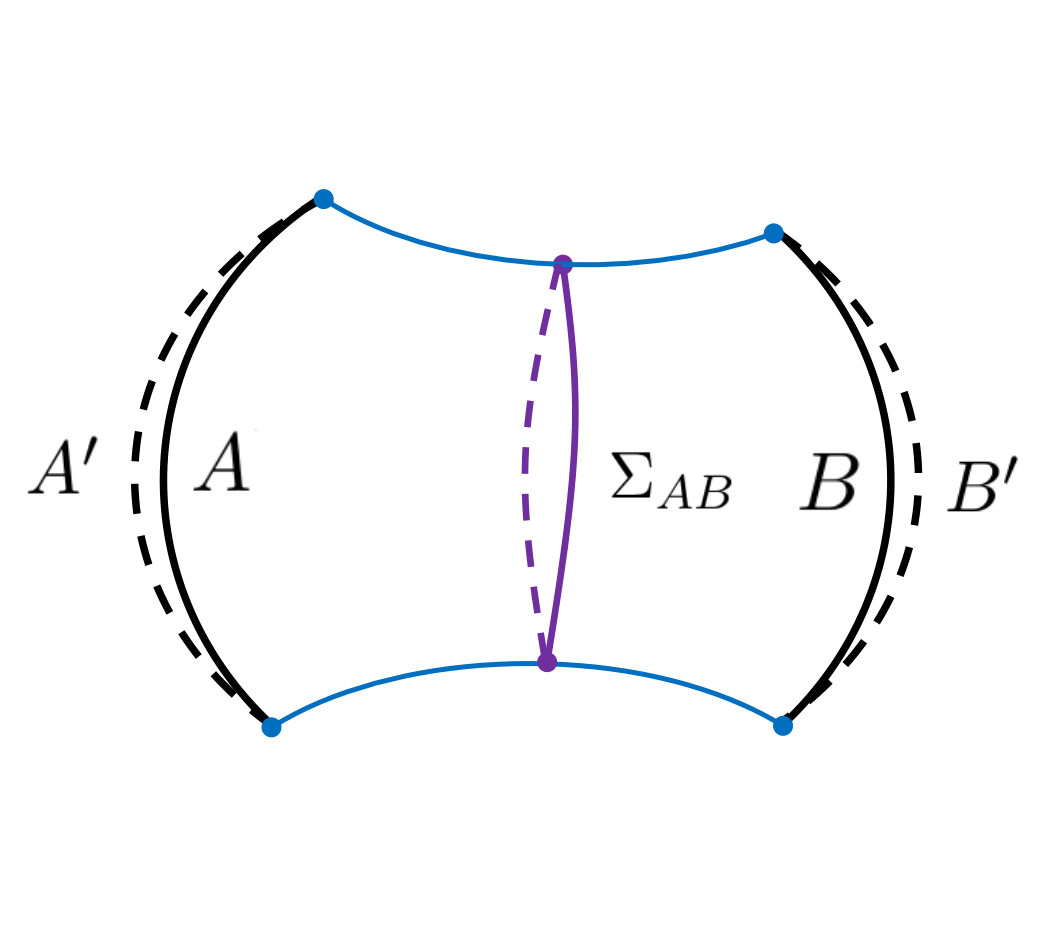}
 \includegraphics[width=0.45\textwidth]{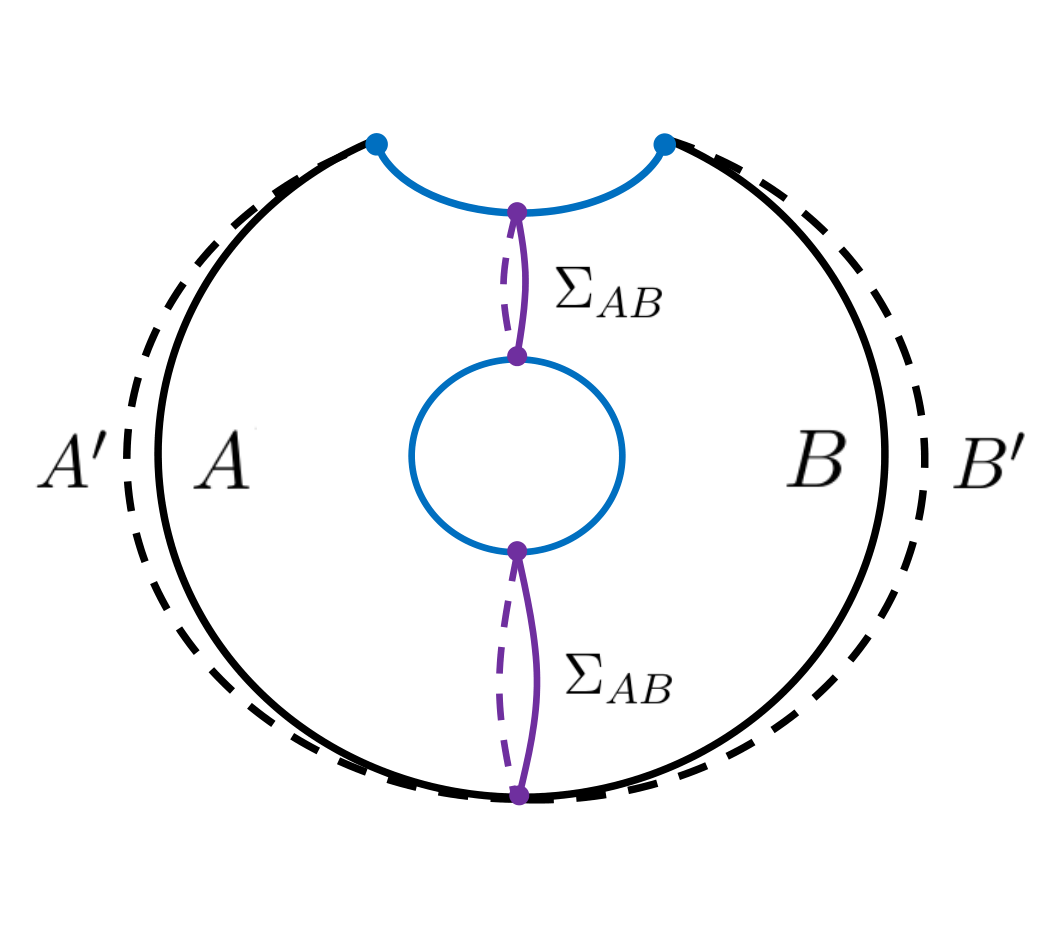}
 \caption{In the left figure $A$ and $B$ are non-adjacent and $\mathcal{W}_{AB}$ is connected. In the right figure there is a black hole in the bulk and $\mathcal{E}_{AB}$ is disconnected. Here the PEE proposal does not apply because $AA'$ is a circle. 
\label{peeandeop5} }
\end{figure}

Then we consider the case where $AB$ is a disconnected but has a connected entanglement wedge, which is shown in the left figure of Fig.\ref{peeandeop5}.  In this case $AA'$ has no boundary hence $\mathcal{E}_{AA'}$ is determined totally by the homology constraint, which turns out to be the minimal circle that warps on the bulk wormhole geometry. It is easy to see the relation \eqref{REMSC} also holds. Note that, in this case the order in $AA'$ is ambiguous hence the PEE proposal does not apply. Using the reflection symmetry, we can easily find that $s_{AA'}(A)=s_{BB'}(A')=\frac{1}{2}S_{R}(A,B)$, hence \eqref{epabsr} follows. 
In the right figure of Fig.\ref{peeandeop5} where $\mathcal{E}_{AB}$ is disconnected, one can also find the relation \eqref{epabsr} holds using similar arguments.


The canonical purification is different from case of global AdS$_3$. It is interesting to find that in both of the two purifications the BPE$(A,B)$ gives the area of $\Sigma_{AB}$, which indicates that the crossing PEE are the same for these two purifications. One can easily check this for the case of Fig.\ref{peeandeop4}, where the crossing PEE can be calculated by the PEE proposal,
\begin{align}
\mathcal{I}_{AB'}=\mathcal{I}_{A'B}=&\frac{1}{2}(S_{AB}+S_{AA'}-S_B-S_{A'})
\cr
=&\frac{c}{6}\log \left(\frac{\sin [(l_{A}+l_B)\pi/L] }{\sin [l_{A}\pi/L]\sin [l_B\pi/L]}\frac{2\sin[l_{A}\pi/L]\sin[l_B\pi/L]}{\sin[(l_A+l_B)\pi/L]}\right)
\cr
=&\frac{c}{6}\log 2\,.
\end{align}
In the above equation we used the relation $S_{A'}=S_A$. Also $S_{AA'}$ is calculated by twice of the length of $\Sigma_{AB}$, which is given by \eqref{adjacentepab}. Again we arrive at the constant $\frac{c}{6}\log 2$ which is the exactly the one we got for global AdS$_3$.

\section{Interpretation for the correspondence between the BPE and the entanglement wedge cross section}\label{section6}
In this section we show that the EWCS can be interpreted as certain types of PEE following the holographic picture for entanglement contour proposed in \cite{Wen:2018whg,Wen:2019ubu}. We find that, the PEE that correspond to $\Sigma_{AB}$ satisfies the balance requirement thus is a BPE. This justifies our previous claim that the BPE gives $\Sigma_{AB}$ when $\rho_{AB}$ has a geometric dual.

\subsection{Brief review on holographic entanglement contour}
\begin{figure}[h] 
   \centering
   \includegraphics[width=0.4\textwidth]{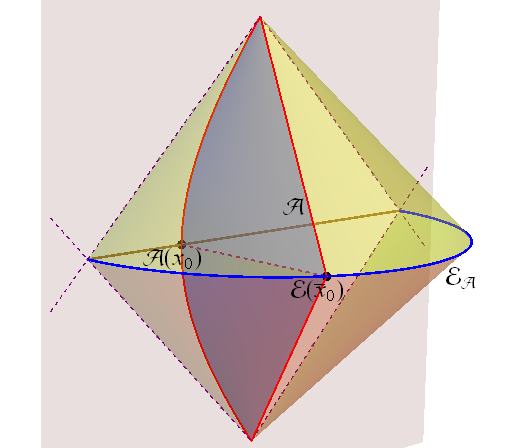}\quad
    \includegraphics[width=0.5\textwidth]{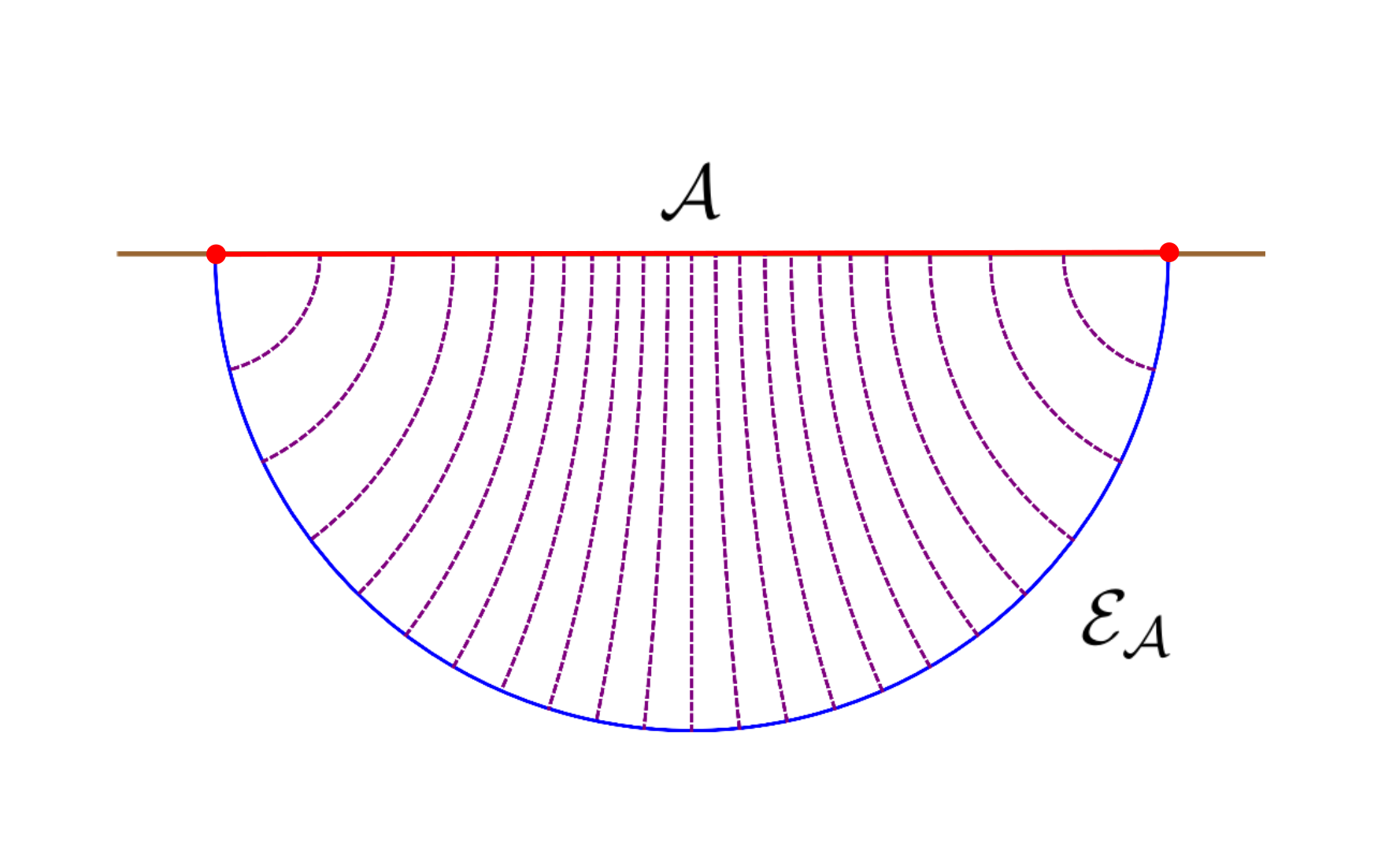}
 \caption{The left figure shows the slicing of the entanglement wedge with modular slices (the blue surface), while the left figure shows a time slice of the entanglement wedge and the fine correspondence between the points in the interval $A$ and its RT surface $\mathcal{E}_{A}$ using the dashed purple geodesics normal to $\mathcal{E}_{A}$.
\label{1} }
\end{figure}

In \cite{Wen:2018whg}, the author gave a holographic picture for the entanglement contour for a single interval in the context of AdS$_3$/CFT$_2$.
Following the bulk and boundary modular flows, it was shown in \cite{Wen:2018whg} that the entanglement wedge $\mathcal{W}_A$ has a natural slicing using the modular slices (see the left figure in Fig.\ref{1}). A modular slice is the orbit of a boundary modular flow line in the bulk\footnote{The modular flow that exactly settled at the boundary has no orbit in the bulk, because the boundary modular flow equals to the bulk modular flow. Here the boundary modular flow is not exactly at but infinitely close to the boundary \cite{Wen:2018whg,Wen:2019ubu}.}. This slicing gives a one-to-one correspondence between the points in the interval $A$ and the points in its RT surface $\mathcal{E}_A$. More explicitly the correspondence means the contribution to $S_A$ from any point in $A$ is represented by its partner point on $\mathcal{E}_A$. 
In all the cases where both of the above geometric construction and the PEE proposal applies, the two proposal give the same results. This consistency even goes beyond the AdS/CFT correspondence \cite{Wen:2018mev}. 

When the interval $A$ is static, the point-to-point correspondence have a simple description using geodesics normal to $\mathcal{E}_{A}$. It was shown in \cite{Han:2019scu} that, any point on $\mathcal{E}_A$ can be connected to its partner point on $A$ via a static geodesic that is normal to $\mathcal{E}_A$ (see the right figure in Fig.\ref{1}). These normal geodesics are where the modular slices intersect with the static homology surface $\mathcal{R}_{A}$.

\begin{figure}[h] 
   \centering
   \includegraphics[width=0.55\textwidth]{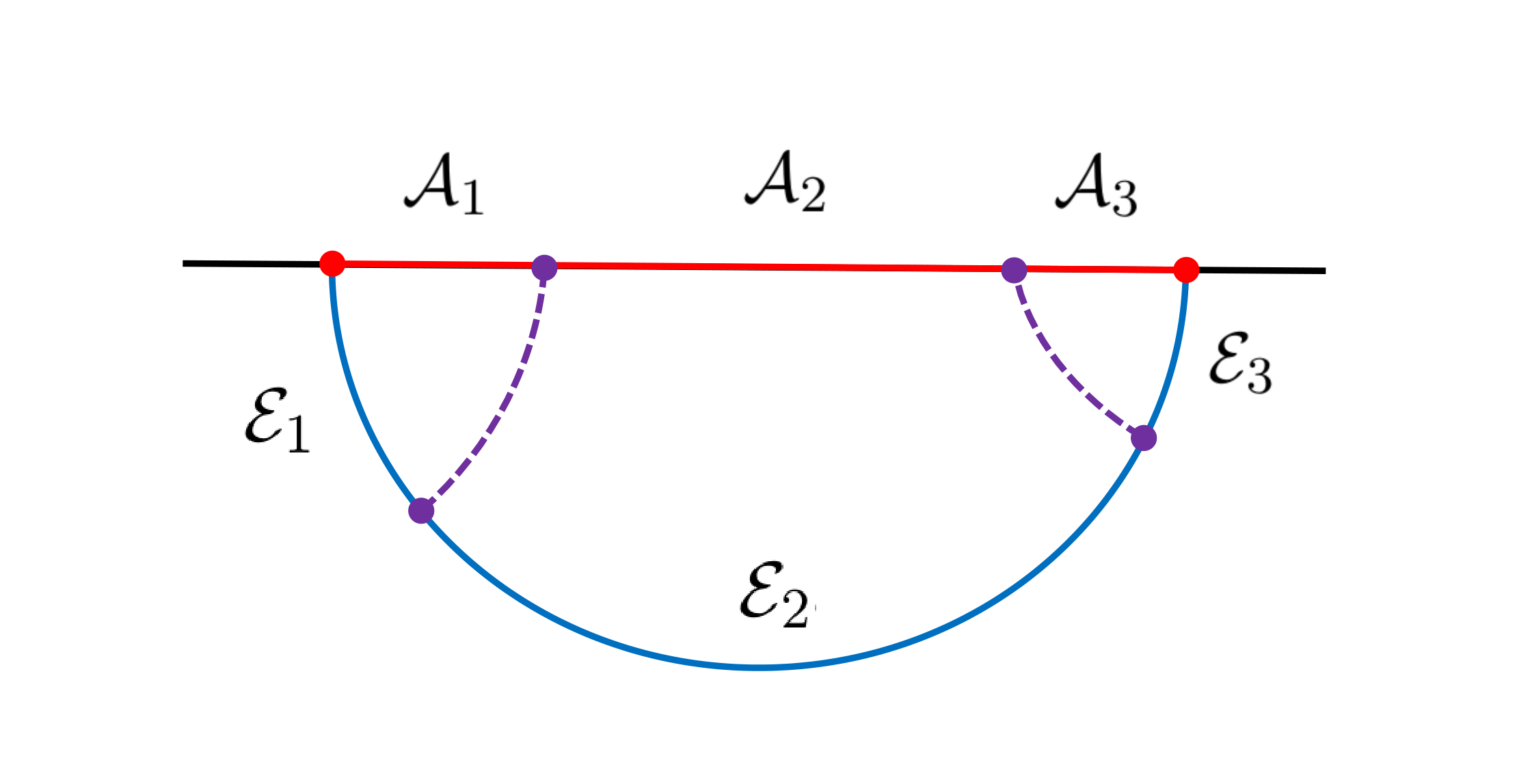} 
 \caption{The purple dashed lines are the static geodesics that normal to $\mathcal{E}_{A}$. They are the lines which set the subsets of $A$ and the subsets of $\mathcal{E}_{A}$ in pairs in the sense of \eqref{pechord}. The points in $A_i$ correspond to the points on $\mathcal{E}_i$.
\label{fig5} }
\end{figure}

In the same sense this correspondence induces a correspondence between the geodesic chords $\mathcal{E}_i$ on $\mathcal{E}_A$ and the PEE of certain subset $A_i$ in $A$,
\begin{align}\label{pechord}
 s_{A}(A_i)=\frac{Length\left(\mathcal{E}_i\right)}{4G}\,.
 \end{align} 
For example see Fig.\ref{fig5}. The above relation gives a finer correspondence between quantum entanglement and bulk geometry than the RT formula.

\subsection{The holographic BPE and the entanglement wedge cross section}

The geodesics normal to the RT surfaces play an important role in the fine correspondence. As we know the EWCS $\Sigma_{AB}$ is also a geodesic chord normal to the RT surface $\mathcal{E}_{AB}$, $\Sigma_{AB}$ can be interpreted as the gravity dual of certain PEE following \eqref{pechord}. In other words, we can relate the length of $\Sigma_{AB}$ to a PEE, and furthermore to a linear combination of the entanglement entropies of relevant boundary intervals following the PEE proposal \eqref{ECproposal}. The prescription to determine this PEE is to extend $\Sigma_{AB}$ to a RT surface of some boundary region. As a portion of this RT surface, $\Sigma_{AB}$ will correspond to a PEE in this region.

\begin{figure}[h] 
   \centering
         \includegraphics[width=0.47\textwidth]{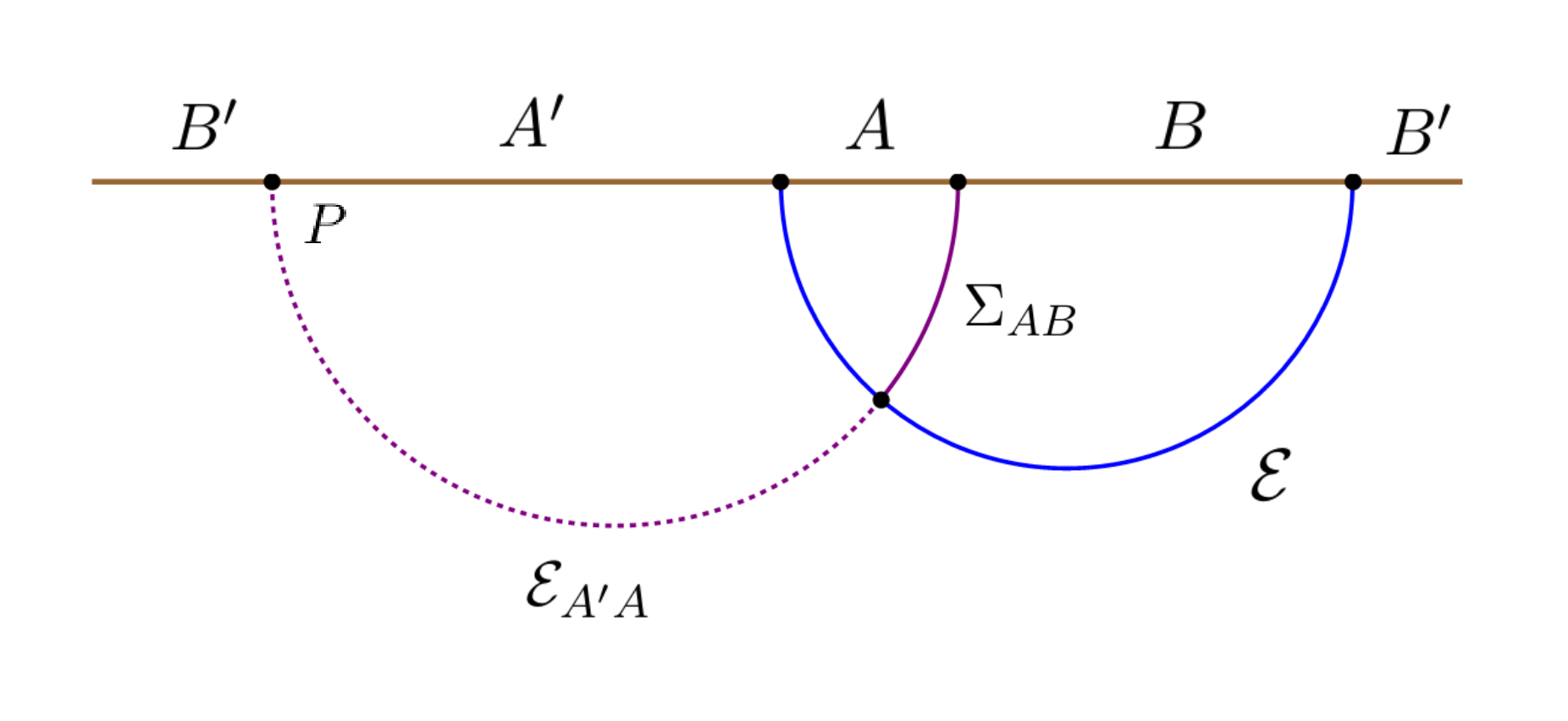}
      \includegraphics[width=0.45\textwidth]{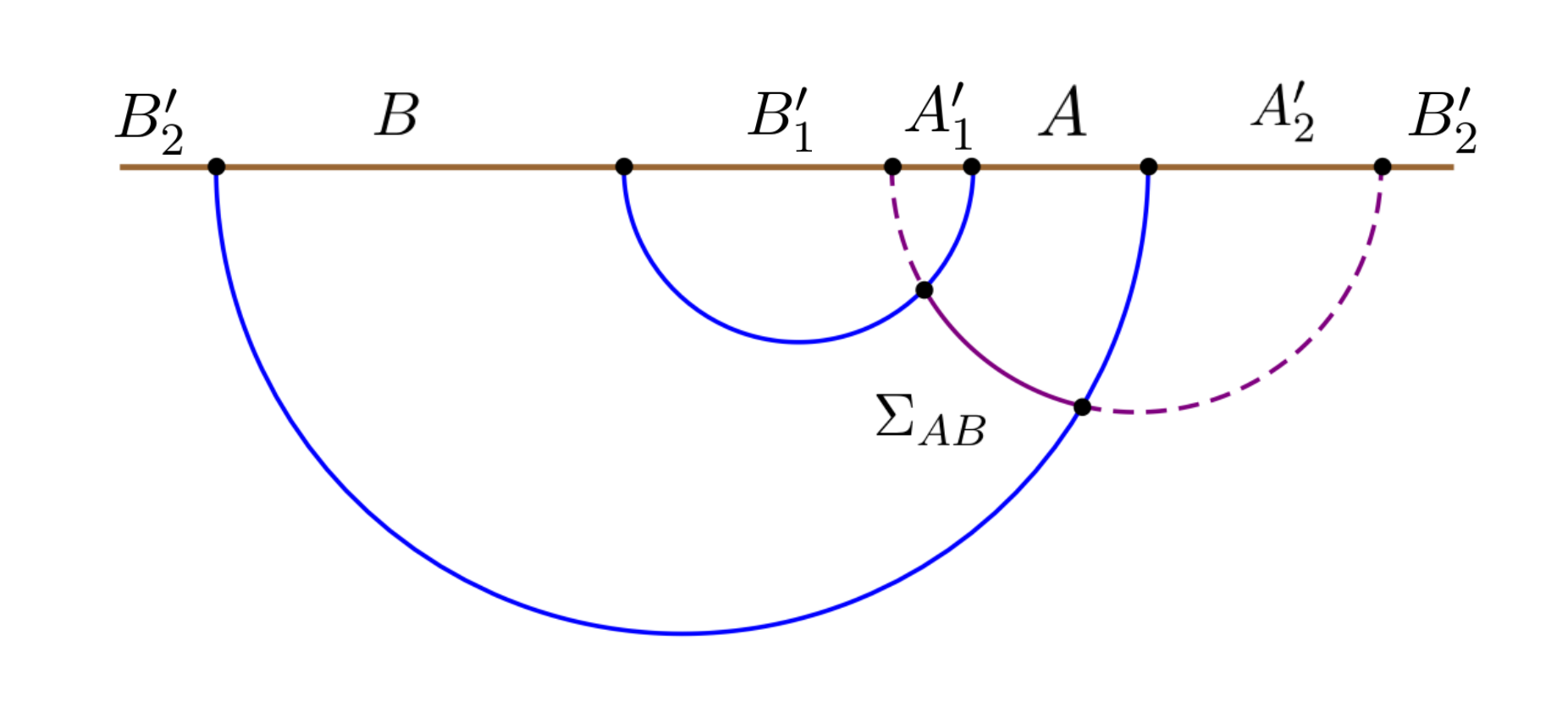}
               \includegraphics[width=0.38\textwidth]{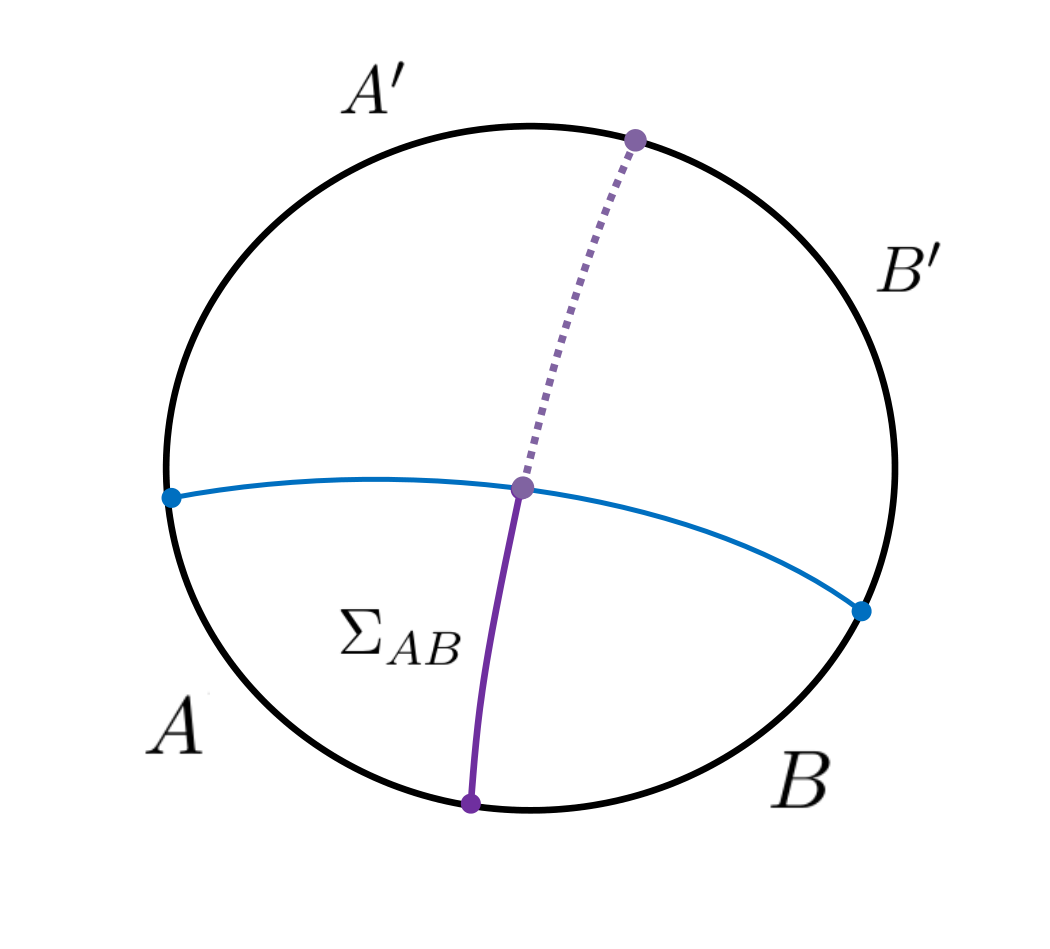} \qquad
   \includegraphics[width=0.38\textwidth]{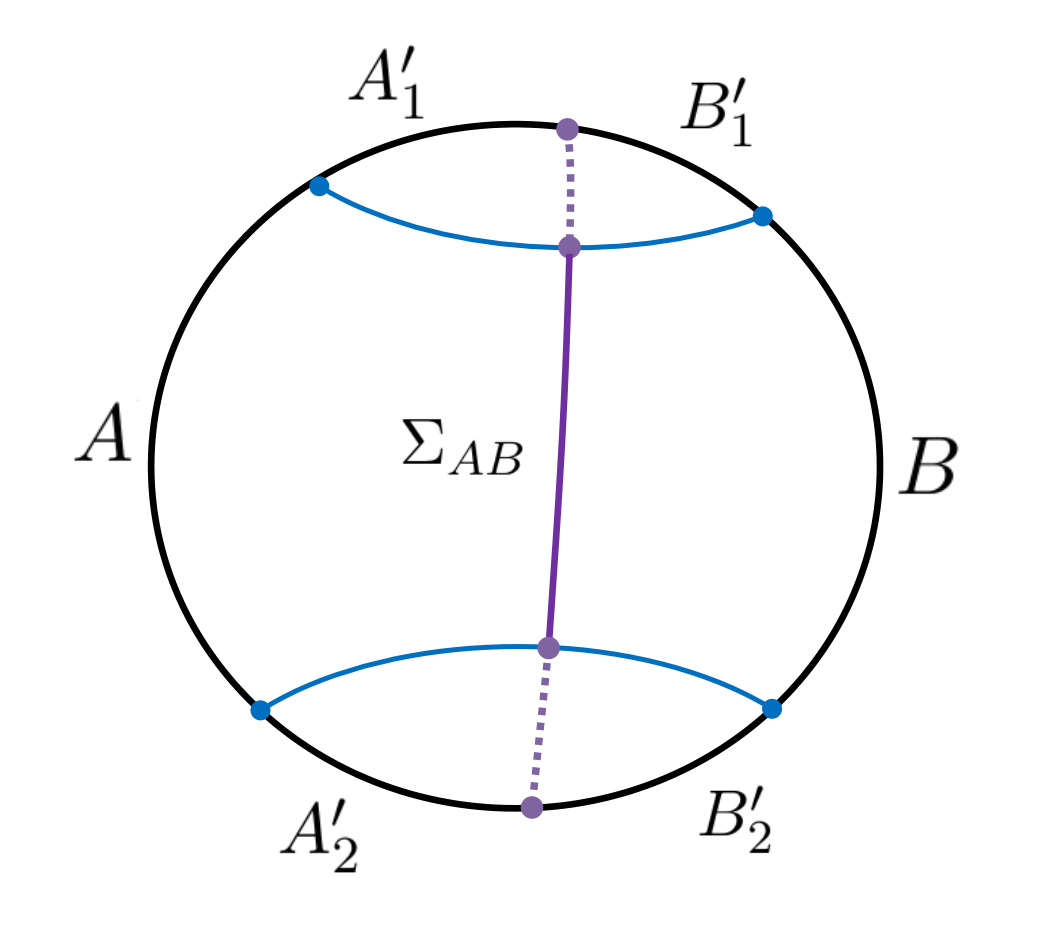} 
 \caption{Here the $\Sigma_{AB}$ is extended to a geodesic that anchored on the boundary, which is just the RT surface $\mathcal{E}_{AA'}$. Then we consider the PEE in $AA'$ and use $\mathcal{E}_{AB}$, which is the blue line normal to $\mathcal{E}_{AA'}$, to 
\label{peeandeop1} }
\end{figure}

Firstly, let us consider the cases in Fig.\ref{peeandeop1} in the context of AdS$_3$/CFT$_2$. The upper figures are Poincar\'e AdS$_3$ and the lower figures are global AdS$_3$. Firstly we determine the RT surface that contains $\Sigma_{AB}$. This can be easily done by extending $\Sigma_{AB}$ to a geodesic anchored on the boundary, which is a RT surface of certain boundary interval. For the left figures in Fig.\ref{peeandeop1}, $AB$ is connected and one of the endpoints of $\Sigma_{AB}$ is settled on the boundary. The extension of $\Sigma_{AB}$ will also intersect with the boundary on another $P$. The point $P$ partition the complement of $AB$ into two parts, which we denote as $A'$ and $B'$. The extension of $\Sigma_{AB}$ is just the RT surface $\mathcal{E}_{AA'}$ (or $\mathcal{E}_{BB'}$) of the interval $A'A$ (or $BB'$). Secondly, since the two RT surfaces $\mathcal{E}_{AB}$ and $\mathcal{E}_{A'A}$ are normal to each other, $\mathcal{E}_{AB}$ can be considered to play the role of the dashed purple line in Fig.\ref{peeandeop1}. According to the fine correspondence \eqref{pechord} we learn that $\Sigma_{AB}$ corresponds to the $s_{AA'}(A)$,
\begin{align}\label{peeeopproposal}
\frac{Area(\Sigma_{AB})}{4G}=s_{A'A}(A)\,.
\end{align}
Note that $\mathcal{E}_{AA'}$ is also the RT surface $\mathcal{E}_{BB'}$ of $BB'$. Using the fine correspondence between the points on $\mathcal{E}_{BB'}$ and $BB'$, we also have
\begin{align}\label{peeeopproposal}
\frac{Area(\Sigma_{AB})}{4G}=s_{BB'}(B)\,.
\end{align}
Then we directly find that the balance requirement is satisfied by the partition induced by the extension of $\Sigma_{AB}$,
\begin{align}
s_{A'A}(A)=s_{BB'}(B)=\text{BPE}(A,B)\,.
\end{align}
One can also explicitly check that the partition of $A'B'$ induced by the balance requirement and the extension of $\Sigma_{AB}$ are exactly the same. This confirms our previous observation that
\begin{align}\label{bpesigmaab}
\text{BPE}(A,B)=\frac{Area(\Sigma_{AB})}{4G}\,.
\end{align}

Similarly, for the non-adjacent $AB$ with connected $\mathcal{W}_{AB}$ (see the figures on the right hand side), the extension of $\Sigma_{AB}$ is a geodesic anchored on the boundary at two points $P_1$ and $P_2$. We see that $P_{1,2}$ partition the boundary into two regions $AA'$ and $BB'$ and the extended $\Sigma_{AB}$ is the RT surface $\mathcal{E}_{AA'}$ of $AA'$. In this case $A'=A'_{1}\cup A'_2$ and $B'=B'_{1}\cup B'_{2}$ are disconnected. Since $\mathcal{E}_{AA'}$ is normal to $\mathcal{E}_{AB}$, according to the fine correspondence \eqref{pechord} and the PEE proposal we have
\begin{align}
 \frac{Area(\Sigma_{AB})}{4G}&=s_{AA'}(A)
 \end{align} 
Also note that $\mathcal{E}_{AA'}=\mathcal{E}_{BB'}$, the fine correspondence between $BB'$ and $\mathcal{E}_{BB'}$ indicates that
\begin{align}
\frac{Area(\Sigma_{AB})}{4G}&=s_{BB'}(B)\,,
\end{align}
hence, again we arrive at \eqref{bpesigmaab}.

Then we discuss the case where the boundary is in a mixed state. For example a thermal state that duals to a BTZ black hole. For simplicity we take $AB$ to be the whole boundary. Here we consider the thermal field double state \cite{Maldacena:2001kr} which is a canonical purification discussed in \cite{Dutta:2019gen}. The gravity dual of the thermofield double state is the eternal black hole \cite{Maldacena:2001kr}. The two copies of CFT on each boundary are entangled thus purifies each other. The Hilbert space factorizes into the left and right subspaces $\mathcal{H}=\mathcal{H}_L\times\mathcal{H}_R$. The thermo-field double state is in the following:
\begin{align}
\ket{\Psi}=\frac{1}{Z}\sum\limits_{n}e^{\frac{-\beta E_n}{2}}\ket{E_n}_L\ket{E_n}_R.
\end{align}
where $E_n$ is the energy eigenvalue of the energy eigenstates $\ket{E_n}$.

\begin{figure}[h] 
   \centering
         \includegraphics[width=0.40\textwidth]{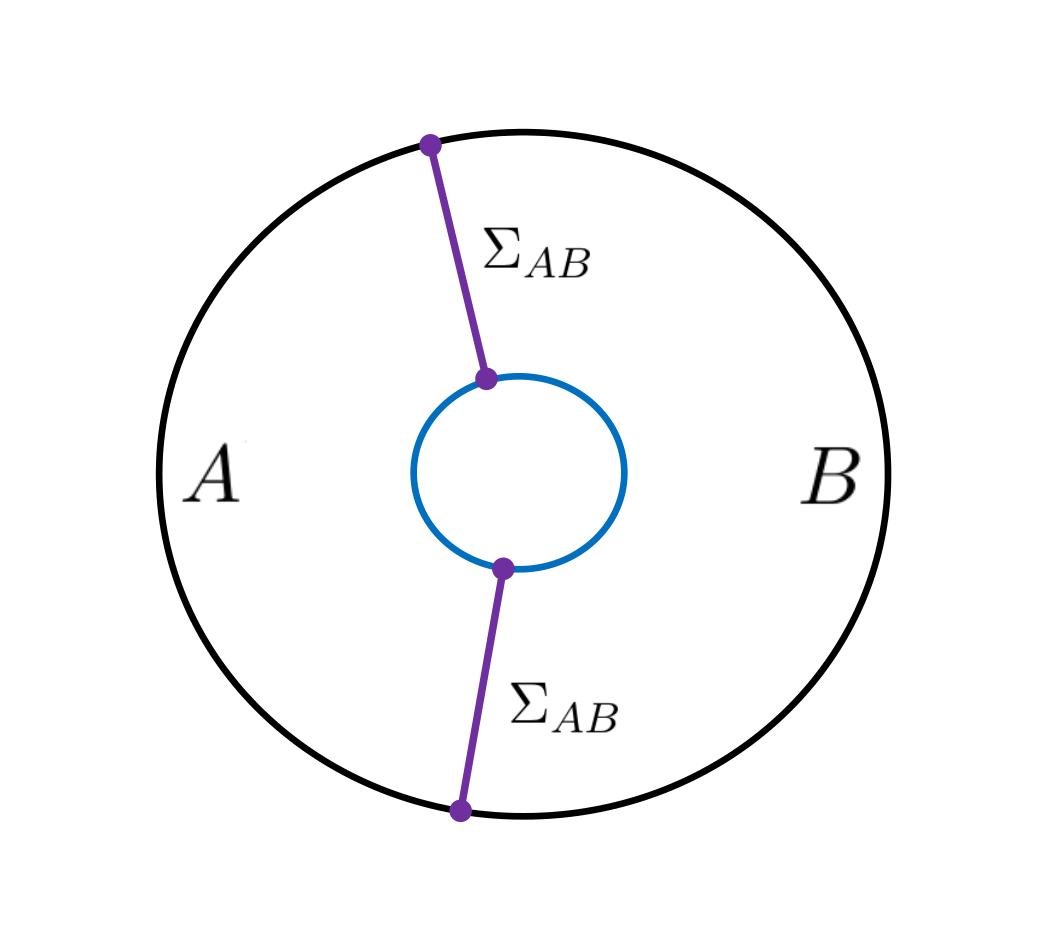}
      \includegraphics[width=0.40\textwidth]{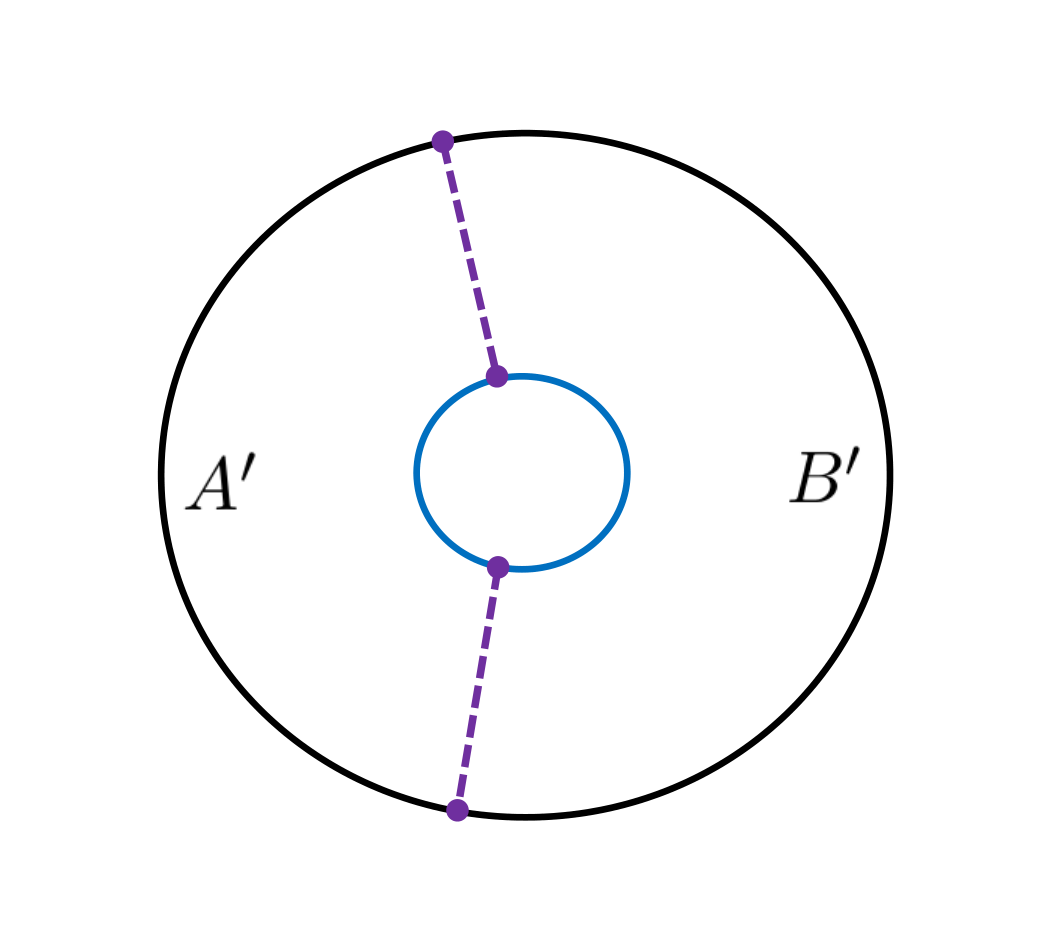}
 \caption{A time slice of the eternal black hole. The blue circle is the horizon, while the purple lines are $\Sigma_{AB}$ and its extension.
\label{peeandeop2} }
\end{figure}

The two figures in Fig.\ref{peeandeop2} draw the left and right hand side of the eternal black hole in a time slice. They are glued together at the horizon. Assuming that $B$ is larger than $A$. When $A$ is small enough the $\Sigma_{AB}$ is just the RT surface $\mathcal{E}_{A}$. As $A$ becomes larger there will be a phase transition for $\Sigma_{AB}$ from $\mathcal{E}_{A}$ to the two geodesic chords emanating from the endpoints of $A$ and intersecting with the horizon vertically (see the two solid purple lines in Fig.\ref{peeandeop2}). Let us consider the later case and use the previous prescription to determine the partition of $A'B'$. Since the two bulk sides are glued together at the horizon, the extended $\Sigma_{AB}$ will enter the right bulk side through the horizon and eventually intersect with the right boundary. See the dashed purple line in Fig.\ref{peeandeop2}. The intersection points are the partition points that divide the right boundary into $A'\cup B'$. The partition determined by the extension of $\Sigma_{AB}$ is exactly a reflection of the partition of $AB$, which is just same as the canonical purification cases. Following our discussion for the canonical purifications, we have
\begin{align}
\text{BPE}(A,B)=\frac{ Area(\Sigma_{AB})}{4G}\,.
\end{align}
When $A$ is small, $\Sigma_{AB}=\mathcal{E}_{A}$ which cannot be extended to the right bulk side. We may need to use the reflection image of $\Sigma_{AB}$ to partition $A'B'$, which is identical to the canonical purifications cases. 

For the cases where $AB$ does not cover the entire boundary, the reflection symmetry between $AB$ and $A'B'$ no longer exist, which differs from the canonical purifications. Since there are subtleties for evaluating the PEEs, we will not discuss for these cases further.

\section{Discussion}\label{sectiond}

The entanglement contour is a finer and more comprehensive description for the entanglement structure of a quantum system. It has the key property of additivity, which makes it different from all the other known entanglement measures. It is natural to expect that, other entanglement measures can be extracted from the entanglement contour. In this paper, we consider a special PEE satisfying the balance requirements for any purification of a bipartite mixed state $\rho_{AB}$. We call it the \textit{balanced partial entanglement} BPE$(A,B,\psi)$ for the purification $\ket{\psi}$, which is omitted when the purification is specified. We find that, for canonical purifications the BPE$(A,B)$ is identical to half of the reflected entropy. While for the holographic purification on the AdS boundary, the BPE$(A,B)$ gives the area of the EWCS divided by $4G$. These results show that, the BPE unifies the quantum information interpretation for the EWCS in both the canonical purification and purifications on the AdS boundaries. Since the BPE can be defined in general quantum systems, in some sense it generalizes the concept of the reflected entropy to generic purifications, and generalize the EWCS to purifications with no geometric description.

Again, note that the partition that satisfies the balanced requirements is not unique. We eliminate this ambiguity by imposing the minimal requirement, which can be satisfied by our prescription to partition the purifier introduced in section \ref{section2}. For continuous systems, the balance requirements can always be satisfied by continuously adjusting the partition. However, for discrete systems, especially few-body systems, the continuous adjusting no longer exist and the number of partitions is finite. In these cases, there is no obvious reason for the existence of a partition that satisfies the balanced requirements. We hope to clarify this point in the future. The study of the BPE in few-body systems is feasible and will be interesting. Calculations in section \ref{section3} and \ref{section4} can also be generalized to lattice models on one-dimensional chains or circles, if the entanglement entropies for single intervals can be calculated.

Also the study of the BPE can be extended to higher dimensions at least for several highly symmetric configurations where the PEE can be explicitly calculated. It will be very interesting to test the relation between the BPE and the EWCS in higher dimensions. We can also explore the relation between the BPE and other entanglement measures like EoP, entanglement negativity and odd entropy in non-holographic systems.

Though the BPE depends on purification, it is independent from a large class of unitary transformations on the complement $A'B'$. The purifications with the same crossing PEE $\mathcal{I}_{AB'}$ gives the same BPE. Interestingly we show that, in both of the canonical purifications and the holographic purification on the boundary of global AdS$_3$, the crossing PEE equals to $\frac{c}{6}\log 2$, which is a constant independent from the partition of the pure state. While for the ``minimal'' purification in the context of the \textit{\textit{surface/state correspondence}}, the crossing PEE equals to $\frac{c}{12}\log 2$. It seems that the crossing PEE is a constant that characterize the purifications, hence could be a useful tool to classify purifications or quantum states. Is there any bounds for the crossing PEE and how can they be saturated? Is the crossing PEE a useful tool to distinguish between holographic and non-holographic states? The physical meaning of these constants deserves further investigation.

The study of BPE bases on our understanding of the PEE or entanglement contour. However, our understanding of the entanglement contour or the PEE is still on a primitive stage. The fundamental definition for the PEE based on density matrix is still not clear and the proposals are not enough to calculate the PEE for even a generic quantum system in two dimensions. It is also important to point out that the BPE is not sensitive to the phase transition between connected and disconnected entanglement wedge. More explicitly the BPE$(A,B)$ does not vanish when the entanglement wedge of $AB$ become disconnected. A naive explanation is that the subset entanglement entropies in the PEE proposal are all connected regions, which are insensitive to the phase transition. This confusion may be understand if we have a deeper understanding of the entanglement contour for disconnected regions.

\section*{Acknowledgments}
The author would like to thank Tatsuma Nishioka, Tadashi Takayanagi and Huajia Wang for helpful discussions. Especially I would like to thank Muxin Han for early collaboration. I thank the Yukawa Institute for Theoretical Physics at Kyoto University. Discussions during the workshop YITP-T-19-03 ``Quantum Information and String Theory 2019'' were useful to complete this work. This work is supported by the NSFC Grant No.11805109 and the ``Zhishan'' Scholars Programs of Southeast University.

\bibliographystyle{JHEP}
 \bibliography{lmbib}

\end{document}